\documentclass[prl,aps,superscriptaddress,footinbib,twocolumn]{revtex4-1}

\usepackage{mathrsfs}
\usepackage{amsfonts}
\usepackage{amssymb}
\usepackage{amsmath}
\usepackage{graphicx}
\usepackage[usenames,dvipsnames]{color}
\usepackage[plainpages=false,colorlinks=true,citecolor=blue,linkcolor=magenta]{hyperref}
\usepackage{lmodern}
\usepackage{amsthm}
\usepackage{dsfont}
\usepackage{natbib}

\AtBeginDocument{%
    \newwrite\bibnotes
    \def\bibnotesext{Notes.bib}
    \immediate\openout\bibnotes=\jobname\bibnotesext
    \immediate\write\bibnotes{@CONTROL{REVTEX41Control}}
    \immediate\write\bibnotes{@CONTROL{%
    apsrev41Control,author="08",pages="1",editor="1",title="1",year="1"}}
     \if@filesw
     \immediate\write\@auxout{\string\citation{apsrev41Control}}%
    \fi
}%

\newcommand{\figpath}{./figures}

\begin{document}

\title{Synthesizing five-body interaction in a superconducting quantum circuit}

\affiliation{Interdisciplinary Center for Quantum Information, State Key Laboratory  of Modern Optical Instrumentation,
and Zhejiang Province Key Laboratory of Quantum Technology and Device,\\
Department of Physics, Zhejiang University, Hangzhou 310027, China\\}
\affiliation{Hangzhou Global Scientific and Technological Innovation Center, Zhejiang University, Hangzhou 311215, China\\}
\affiliation{CAS Center of Excellence in Topological Quantum Computation, Beijing
100190, China\\}
\affiliation{Institute of Quantum Science and Engineering, Texas A$\&$M University,
College Station, Texas 77843, USA}

\author{Ke Zhang$^{1, *}$}
\author{Hekang Li$^{1, *}$}
\author{Pengfei Zhang$^{1}$}
\author{Jiale Yuan$^{1}$}
\author{Jinyan Chen$^{1}$}
\author{Wenhui Ren$^{1}$}
\author{Zhen Wang$^{1}$}
\author{Chao Song$^{1, \dagger}$}
\author{Da-Wei~Wang$^{1,3, \ddagger}$}
\author{H. Wang$^{1,2}$}
\author{Shiyao Zhu$^{1,2}$}
\author{Girish S. Agarwal$^{4}$}
\author{Marlan O. Scully$^{4, \S}$}
\begin{abstract}
Synthesizing many-body interaction Hamiltonian is a central task in quantum simulation. However, it is challenging to synthesize interactions including more than two spins. Borrowing tools from quantum optics, we synthesize five-body spin-exchange interaction in a superconducting quantum circuit by simultaneously exciting four independent qubits with time-energy correlated photon quadruples generated from a qudit. During the dynamic evolution of the five-body interaction, a Greenberger-Horne-Zeilinger state is generated in a single step with fidelity estimated to be $0.685$. We compare the influence of noise on the three-, four- and five-body interaction as a step toward answering the question on the quantum origin of chiral molecules. We also demonstrate a many-body Mach-Zehnder interferometer which potentially has a Heisenberg-limit sensitivity. This study paves a way for quantum simulation involving many-body interactions and high excited states of quantum circuits.
\end{abstract}

\maketitle

The synthesis of many-body interaction Hamiltonian plays a vital role in quantum simulation and quantum computing. Most quantum gates \cite{Barenco1995} rely on two-body interactions, based on which state-of-the-art quantum circuits have been built \cite{Arute2019, Jurcevic2020, Zhu2021} and quantum supremacy has been claimed \cite{Arute2019,Zhu2021}. However, to exploit the full degree of freedom in simulating emergent many-body physics with superconducting circuits, we need to synthesize arbitrary interaction between qubits \cite{Babbush2018,Bharti2021}, although quantum algorithms can partially help \cite{Peng2009,Georgescu2014}. The anti-symmetric spin-exchange interaction (Dzyaloshinskii-Moriya interaction) \cite{Dzy1957,Moriya1960} has been synthesized by breaking the time-reversal symmetry through Floquet modulation \cite{Wang2019}, enabling a three-qubit chiral quantum gate for entangling qubits more efficiently than the two-qubit gates \cite{Kelly2015}. Similar techniques have been applied to the synthesis of effective gauge field \cite{Roushan2017} and three-spin chirality Hamiltonian \cite{Liu2020}, which is a necessary element in simulating chiral spin liquid \cite{Wen1989} and promising to realize the topological states of quantized light \cite{Cai2021,Yuan2021,Cheng2021}. Four-spin ring-exchange interaction, which has been synthesized in optical lattices of cold atoms \cite{Dai2017}, is essential for toric codes in topological quantum computing \cite{Kitaev2003}. However, interactions involving more than three spins have never been achieved in superconducting qubits, due to the difficulty in realizing ultrastrong coupling and eliminating lower order interactions \cite{Garziano2016,Kockum2017,Zhao2017,Stassi2017,Liu2019,Munoz2020}. 

Superconducting quantum circuits are a competitive platform for quantum simulation thanks to its remarkable scalability and tunability \cite{You2011,Devoret2013, Kjaergaard2020}. However, the direct capacitive or inductive coupling between superconducting qubits is limited to two-body interactions \cite{Krantz2019}. Recently, a quantum optics technique of entangling non-interacting atoms with time-energy entangled photon pairs \cite{Muthukrishnan2004,Zhang2013,Dorfman2016} is implemented in the strong coupling regime of superconducting circuits, realizing a three-body spin-exchange interaction with high fidelity \cite{Ren2020}. Here we report the synthesis of five-body spin-exchange interaction with four time-frequency correlated photons in a superconducting circuit. By preparing a qudit in the fourth excited state, time-energy correlated photon quadruples are generated in cascade transitions and coupled to four independent qubits, such that all the four qubits are excited simultaneously while the qudit directly transits from the fourth excited state to the ground state. In this process, five-body spin-exchange interaction is synthesized for the first time. Apart from the application in quantum simulation, such multi-spin interactions can be used as a nonlinear interferometer in Heisenberg-limit metrology since it generates multi-spin entanglement in a single step. It can also be used to simulate the quantum tunneling between left- and right-handed molecules, which is a step toward answering the question on the origin of chiral molecules.

\begin{figure}[t]
		\includegraphics[width=\linewidth]{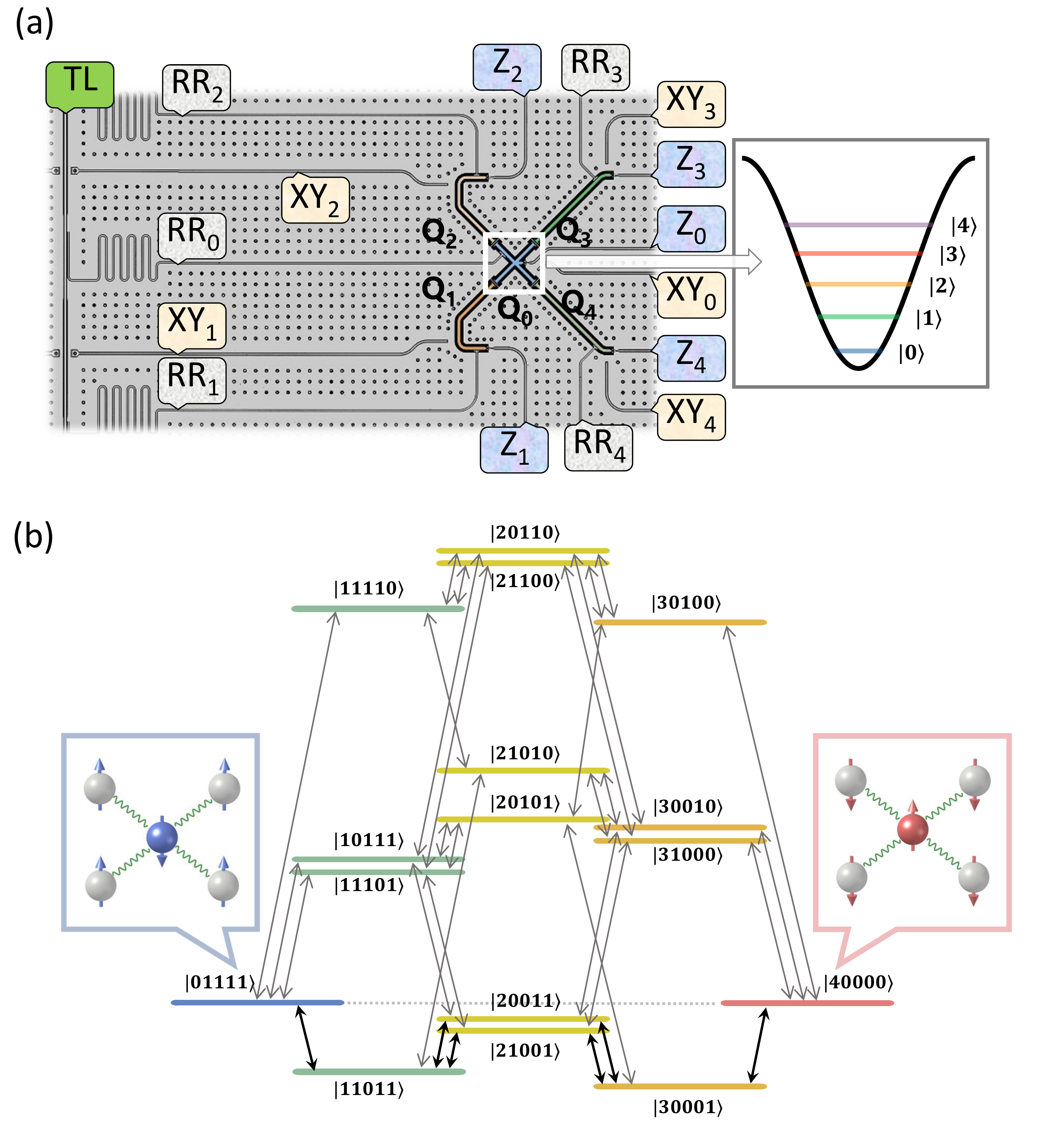}
		\caption{
			(a) Device image illustrating the five frequency-tunable transmon circuits labeled from Q$_0$ to Q$_4$, with Q$_0$ surrounded by Q$_1$ to Q$_4$. Due to the nonlinear Josephson inductance, the energy potential of transmons is sinusoidal (black line in the right panel), which allows us to isolate the $m$ lowest energy levels (colored lines in the right panel) as a qubit (Q$_1$ to Q$_4$ with $m=2$) or qudit (Q$_0$ with $m=5$). Each circuit Q$_j$ has its own flux bias line Z$_j$ for fast frequency tuning, microwave line XY$_j$ for SU(2) spin rotation, and readout resonator RR$_j$ that couples to a common transmission line TL for dispersive readout of Q$_j$'s state. The dots are bumps for the flip-chip process. (b) Energy configurations of the qudit and four qubits for 5-body interaction.
		}
		\label{fig:Device}
\end{figure}

 The experiment is performed in a superconducting circuit where four transmon qubits are symmetrically coupled to a central transmon qudit (see Fig.~\ref{fig:Device}(a)). All the five transmon circuits have a sinusoidal potential well that hosts multiple energy levels~\cite{libo2018}. The four surrounding qubits Q$_j$ with $j=1$ to $4$ are used as two-level artificial atoms, while the central qudit Q$_0$ plays the role of a four-level atom, which generates cascade time-energy correlated photon quadruples. The transition frequency from the ground state $|0\rangle$ to the first excited state $|1\rangle$ of the transmon circuits are tunable from 4 to 6 GHz. The coupling strengths $g_j/2\pi$ between the surrounding qubits Q$_j$ and the central qudit Q$_0$ are around 23 MHz, while those between the surrounding qubits are smaller than 1 MHz (see Supplementary Material \cite{SupMat} for details on the device, operation and readout). The Hamiltonian of the system in the rotating wave approximation is given by (we set $\hbar=1$),
\begin{equation}
\begin{aligned}
H=&\sum_{n=1}^{4}\sum_{k=1}^n\nu_{k}|n_0\rangle\langle n_0| + \sum_{j=1}^{4}\omega_j |1_j\rangle\langle 1_j| \\
&+\sum_{n=1}^4\sum_{j=1}^{4}\sqrt{n}g_j(S_{n}^+\sigma_j^- + \sigma_j^+S_{n}^-),
\end{aligned}
\label{h}
\end{equation}
where $|n_j\rangle$ is the $n$th ($n=0,1,2,...$) level of Q$_j$, $S_n^+\equiv |n_0\rangle\langle (n-1)_0|$ are the raising operators between the adjacent levels of Q$_0$, and $\sigma_j^+\equiv |1_j\rangle\langle 0_j|$ are the raising operators of the qubit Q$_j$ with $S_n^-$ and $\sigma^-$ being their corresponding lowering operators, $\nu_{k}$ is the $k$th transition frequency between the energy levels $|k_0\rangle$ and $|(k-1)_0\rangle$ of the qudit Q$_0$, and $\omega_j$ is the transition frequency of the qubit Q$_j$.

The five-body spin-exchange interaction Hamiltonian can be realized by carefully arranging the transition frequencies of all qubits such that the four-photon resonance is satisfied, $\sum_{k=1}^4\nu_k=\sum_{j=1}^4\omega_j$, while the single photon, two-photon and three-photon resonances are avoided, as shown by the energy level diagram in Fig.~\ref{fig:Device}(b). It is easy to know that when we drive all qubits with uncorrelated photons, even if the four-photon resonance is satisfied, the qubits cannot be simultaneously excited, because all possible quantum paths cancel out (see details in \cite{SupMat}). The key element in our approach is that the qudit emits four photons sequentially with different frequencies $\nu_4$, $\nu_3$, $\nu_2$ and $\nu_1$. The total number of possible quantum paths between the states $|40000\rangle\equiv |4_00_10_20_30_4\rangle$ and $|01111\rangle$ is reduced by a factor of $4!=24$ compared with the case when the qubits are driven by uncorrelated photons. The quantum paths that could have cancelled the remaining $24$ paths are removed by the constraint of the time-energy correlation between the four cascaded photons.
The resulted effective Hamiltonian is,
\begin{equation}
\begin{aligned}
H_\text{eff}=\lambda\Xi_4^-\sigma_1^+\sigma_2^+\sigma_3^+\sigma_4^+ + h.c.,
\end{aligned}
\end{equation}
where $\Xi_n^- \equiv |0_0\rangle\langle n_0|$ is the lowering operator of the qudit from the $n$th excited state to the ground state, and $\lambda$ is the effective coupling strength.

\begin{figure}[h]
		\includegraphics[width=\linewidth]{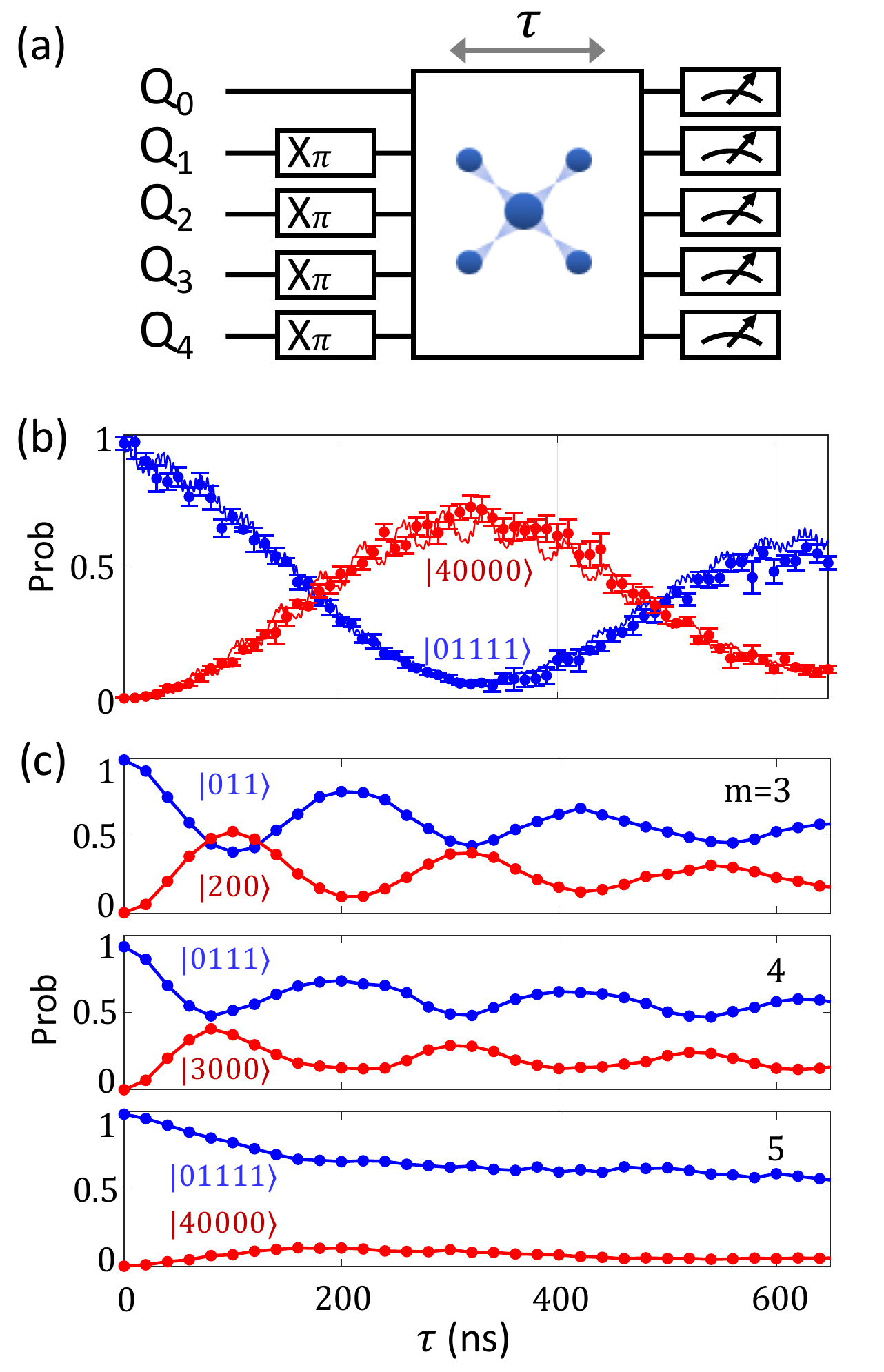}
		\caption{
			(a) Sequence diagram for observing the system's dynamic evolution. After preparing the system to the initial state $|01111\rangle$ by applying $X_\pi$ rotation (a $\pi$ rotation around x axis) to Q$_1$ to Q$_4$, we quickly tune the transition frequencies of each qubit to activate the 5-body interaction. After a specific time $\tau$, the occupational probabilities of the system for different computational state are measured. (b) The experimentally measured occupational probabilities of $|c_1(\tau)|^2$ for $|01111\rangle$ (blue dots) and $|c_2(\tau)|^2$ for $|40000\rangle$ (red dots) for different interaction times $\tau$. Error bars represent statistical errors. Lines are the results obtained by numerical simulation, where 5 and 3 levels are considered for Q$_0$ and other qubits respectively. (c) The effect of noise on the evolution under the $m$-body spin-exchange Hamiltonian. The low-frequency noise is simulated by a random detuning between the two states in the interaction.
		}
		\label{fig:Evolution}
\end{figure}

The controlling sequence diagram for the dynamics of $H_\text{eff}$ is sketched in Fig.~\ref{fig:Evolution}(a). We prepare the qubits in the initial state $|\Psi(0)\rangle = |01111\rangle$ at their idle frequencies. Then the four qubits are quickly biased to their interaction frequencies $\omega_j$. After an interaction time $\tau$, we bring the qubits to their readout frequencies for measurement. The detailed values of transition frequencies for qubit initialization and readout can be found in~\cite{SupMat}. The results of the joint measurement of the wave function $|\Psi(\tau)\rangle = c_1(\tau)|01111\rangle+c_2(\tau)|40000\rangle$, ignoring the insignificant terms, are shown in Fig.~\ref{fig:Evolution}(b), where the experimentally obtained probabilities of $|c_1(\tau)|^2$ and  $|c_2(\tau)|^2$ (colored dots with errorbars) are plotted in comparison with the numerical simulation (colored lines) obtained from the original Hamiltonian in Eq.~(\ref{h}). The Rabi oscillation between the two states is observed as expected. In the numerical simulation, we use the Lindblad master equation with the experimentally measured energy relaxation time $T_{1,j}$ and empirical pure dephasing time $T_{\varphi,j}$ ($\approx 6T_{2,j}^*$ where $T_{2,j}^*$ is the experimentally measured Ramsey Gaussian dephasing time) to capture the impact of decoherence ~\cite{Song2019, Guo2021}.  The values of $T_{1,j}$, $T_{2,j}^*$ are found in \cite{SupMat}. 

The interaction strength of the five-body spin-exchange interaction is estimated to be $\lambda/2\pi\approx 0.8$ MHz, which is strong relative to the decoherence time and allows the generation of a five-spin Greenberger-Horne-Zeilinger (GHZ) state in a single step. At $\tau=170$ ns, the state evolves to $|\Psi\rangle=(|01111\rangle+e^{i\phi}|40000\rangle)/\sqrt{2}$ with $\phi$ being a trivial dynamical phase. The measured fidelity of the experimental data is estimated to be $0.685\pm 0.022$ by directly measuring the four non-zero elements of the density matrix from the many-body interference, which is consistent with the lower bound obtained by quantum state tomography (see Supplementary Material \cite{SupMat}) and satisfies the criterion of global entanglement. In the traditional S{\o}rensen-M{\o}lmer approach \cite{Sorensen2000} of single-step generation of GHZ state of $N$ qubits, all the $N+1$ states in the symmetric subspace are involved, while in our approach only the two relevant states are involved, such that we can simulate the quantum tunneling between the left- and right-handed molecules.

Since early days of quantum mechanics, the origin of chiral molecules has puzzled generations of physicists \cite{Hund1927,Cina1995,Mason1984,Darqui2010,Trost2009,Houck2012}. In particular, Hund argued that the parity operator commutes with the electromagnetic interaction Hamiltonian, which is responsible for the formation of molecules. Therefore, the eigenstates of molecules shall be eigenstates of the parity operator, i.e., a superposition of the left- and right-handed molecular states. As a consequence, chiral molecules are not in their eigenstates and shall not be stable. However, in real world chiral molecules are ubiquitous and stable, while a superposition of two chiral molecules is considered to be a Schr\"odinger cate state, which is unnatural and generally hard to realize. This contradiction between the quantum mechanical prediction and reality is called the Hund paradox  \cite{Hund1927}. 

To resolve the paradox, an argument is that the left- and right-handed molecules reside in two energy valleys and the tunneling strength between them is so small that it takes days (and even lifetime of the universe) for large chiral molecules to tunnel from one configuration to the other. During this process environmental noises induce decoherence and hinder the tunneling, which is similar to the quantum Zeno effect \cite{Trost2009}. 
The two quantum states involved in the five-body interaction can be used to simulate the tunneling between the left- and right-handed molecular states, such that the effect of environmental noises can be investigated toward the question on the stablization of chiral molecules. 

As a step in this direction, we simulate the suppressed tunneling between two chiral molecules due to the slow environmental noises, which can also be considered as a random potential difference between the two chiral molecules. To demonstrate this, we artificially inject arbitrary flux noises to the system during the five-body interaction. Each circuit Q$_j$ is offset from its interaction frequency by a small amount of $\delta_{j,k}$, which is randomly chosen in a range of $[-\Delta_j, \Delta_j]$ but fixed for the $k$th pulse sequence. In the experiment, the noise strength $\Delta_0/2\pi\approx5$ MHz for Q$_0$ between the transition from the state $|0_0\rangle$ to $|4_0\rangle$. The noise range $\Delta_j/2\pi$ is set to be about $5$ MHz for all other qubits Q$_j$. Assuming that the noise is slow, an ensemble of 20 pulse sequences are applied to emulate the random white noise that shifts the energy of Q$_j$. For each sequence, we record the probabilities of the two states as a function of time. We average over 20 traces and find that the oscillation between the two states vanishes, as shown in Fig.~\ref{fig:Evolution}(c). By exciting Q$_0$ to the second or third excited state and coupling it to two or three qubits, we synthesize the three- or four-body spin-exchange Hamiltonian, $\lambda_3 \Xi_2^- \sigma_1^+ \sigma_2^+ +h.c.$ or $\lambda_4 \Xi_3^- \sigma_1^+ \sigma_2^+ \sigma_3^++h.c.$, where $\lambda_3/2\pi\approx2.25$ MHz and $\lambda_4/2\pi\approx 2.29$ MHz are the interaction strengths. The same noise strength has a smaller effect on the four- and three-body interaction, resulting from a larger interaction strength and a weaker noise-induced decoherence effect. The oscillation between the two states are still visible, although partially smeared by the noise, as shown in Fig.~\ref{fig:Evolution}(c).

\begin{figure}[t]
	\begin{center}
		\includegraphics[width=\linewidth]{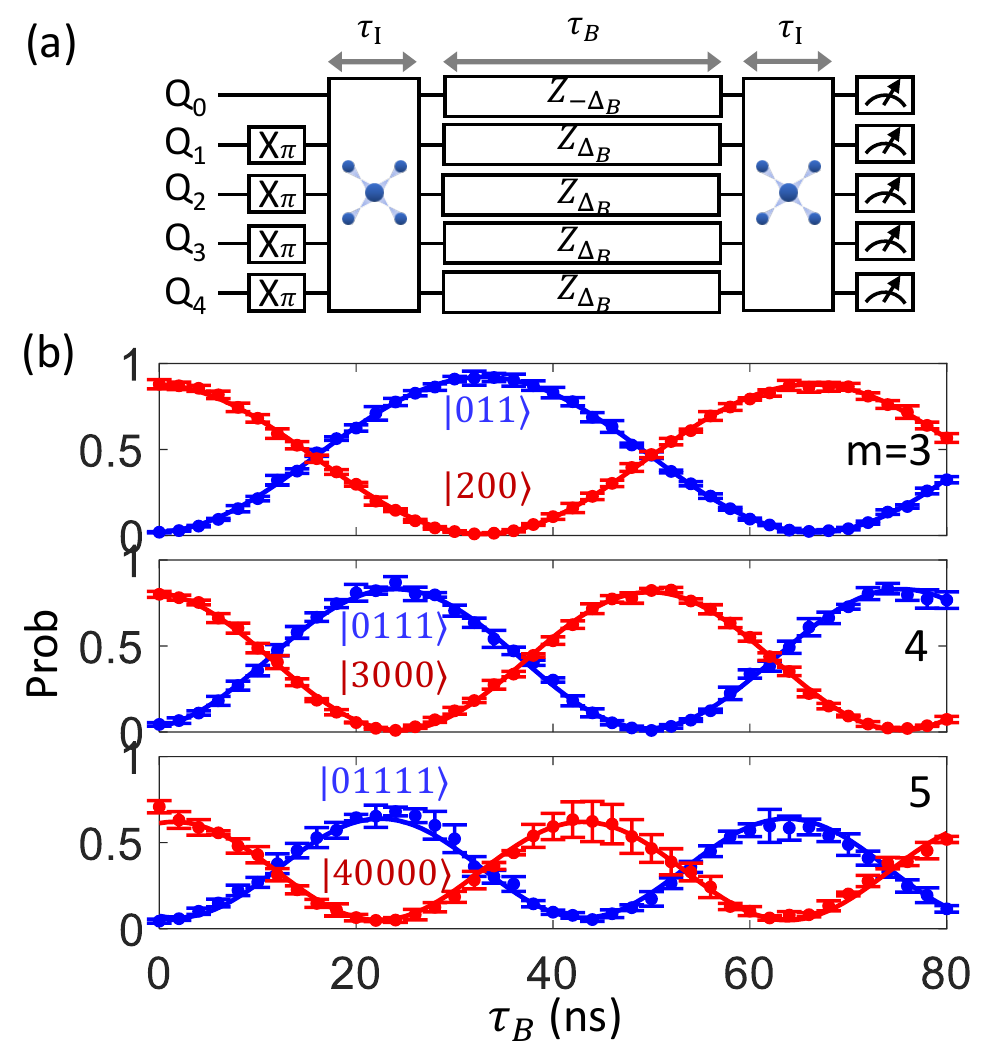}
		\caption{
			(a) Sequence diagram for detecting magnetic field leveraging the $m$-body interaction, with $m=5$ here as an example.
			The magnetic field is synthesized by applying to each transmon circuit a square Z pulse to offset its transition frequency between states $|0\rangle$ and $|m-1\rangle$ ($|1\rangle$) for Q$_0$ (other qubits) by an amount of $-\Delta_B$ ($\Delta_B$) for a time $\tau_B$.
			The dynamical phase accumulated during this process can be detected by sandwiching itself between two $m$-body interaction operations with a fixed time $\tau_{\text{I}} \sim 2\pi/8\lambda_m$.
			(b) The experimentally measured occupational probabilities of $|c_1(\tau_B)|^2$ (blue dots) and $|c_2(\tau_B)|^2$ (red dots) for different time $\tau_B$. Lines are the fitting results. The fitted oscillation frequencies for $m = 3, 4, 5$ are 14.9, 19.5 and 24.2 MHz respectively, agreeing well with the ratio of $3:4:5$.
		}
		\label{fig:Interfrometer}
	\end{center}
\end{figure}

The many-body spin-exchange Hamiltonian can be used to build a Mach-Zehnder interferometer that has a Heisenberg-limit sensitivity \cite{Facchi2011}. In the $m$-body interferometer, we introduce an energy-splitting Hamiltonian $B_z(\sum_{j=1}^{m-1}\sigma^z_j-\Xi_{m-1}^z)$ where $\Xi_m^z\equiv |m_0\rangle\langle m_0|-|0_0\rangle\langle 0_0|$ and $B_z$ is an artificial magnetic field. The dynamic phase induced by this Hamiltonian can be detected as follows. 
We first prepare the state $|0\rangle \otimes |1\rangle^{\otimes (m-1)}$ and then activate the $m$-body interaction for a fixed time $\tau_{\text{I}} \sim 2\pi/8\lambda_m$ to steer the system to the GHZ state. In the experiment, $\tau_{\text{I}}$ is slightly adjusted for an optimized GHZ state fidelity and set to be 60 ns (55 ns, 170 ns) for $m$ = 3 (4, 5).
Then we apply local magnetic field on each spin-1/2 particle, which is synthesized by applying a square Z pulse to each transmon circuit for a specific time $\tau_{B}$ and with a strength of $-\Delta_B$ (between states $|0\rangle$ and $|m-1\rangle$) for Q$_0$ and $\Delta_B$ for other qubits. $\Delta_B/2\pi$ is fixed to be around 5 MHz in the experiment.
After activating the $m$-body interaction again for a time $\tau_{\text{I}}$, we measure the occupational probabilities of $|c_1(\tau_B)|^2$ for $|0\rangle\otimes|1\rangle^{\otimes (m-1)}$ and $|c_2(\tau_B)|^2$ for $|m-1\rangle\otimes|0\rangle^{\otimes (m-1)}$. 
The controlling sequence for a 5-body interferometer is shown in Fig~\ref{fig:Interfrometer}(a). The data is shown in Fig.~\ref{fig:Interfrometer}(b). 
The oscillation frequency scales linearly with $m$. 
Although the sensitivity is not enhanced to the Heisenberg limit due to a lowered visibility of the oscillation for larger $m$, we note that the oscillation amplitude infers the off-diagonal term $\rho_{\text{off}}$ of the GHZ state by $|c_1(\tau_B)|^2 = -|\rho_{\text{off}}|\cos(2\pi m\Delta_B\tau_B)+\text{const}$, which dramatically reduce the number of quantum operations required to benchmark the GHZ state fidelity compared with the traditional tomography method.
We note that the GHZ state obtained from the all-to-all spin interaction can also be used to build a Heisenberg interferometer \cite{Agarwal}. 
However, in that case all states in the symmetric subspace are involved and the approach is more prone to noise compared with the current approach, which only involves two states.

\textit{Conclusion.} ---We propose a method to synthesize multi-body interaction in a superconducting superconducting quantum circuit.
We experimentally realized $m$-body interactions with $m$ up to 5 and investigated the effect of noise on the many-body interactions, which is a step toward finding the quantum origin of chiral molecules. We further built a many-body interferometer and the linear $m$-scaling oscillation frequency of the interference pattern was demonstrated, which is promising for achieving Heisenberg-limited sensitivity. We demonstrated that the fidelity of the GHZ state obtained from such an interferometer has an enhanced precision compared with the conventional tomography method.

\begin{acknowledgments}
	Devices were made at the Micro-Nano Fabrication Center of Zhejiang University. 
	We are grateful to the authors of QuTip for making their package public. We acknowledge the support of 
	the National Natural Science Foundation of China (Grants No. 11725419, U20A2076, 92065204 and 11934011), 
	the National Basic Research Program of China (Grants No. 2017YFA0304300 and 2019YFA0308100),
	the Zhejiang Province Key Research and Development Program (Grant No. 2020C01019), 
	and the Key-Area Research and Development Program of Guangdong Province (Grant No.2020B0303030001).
\end{acknowledgments}

\noindent{$^*$ K. Z. and H.K. L. contributed equally to this work.\\
$^\dagger$ chaosong@zju.edu.cn\\
$^\ddagger$ dwwang@zju.edu.cn\\
$^\S$ scully@tamu.edu}

\bibliography{fiveBody}

\begin{thebibliography}{47}%
\makeatletter
\providecommand \@ifxundefined [1]{%
 \@ifx{#1\undefined}
}%
\providecommand \@ifnum [1]{%
 \ifnum #1\expandafter \@firstoftwo
 \else \expandafter \@secondoftwo
 \fi
}%
\providecommand \@ifx [1]{%
 \ifx #1\expandafter \@firstoftwo
 \else \expandafter \@secondoftwo
 \fi
}%
\providecommand \natexlab [1]{#1}%
\providecommand \enquote  [1]{``#1''}%
\providecommand \bibnamefont  [1]{#1}%
\providecommand \bibfnamefont [1]{#1}%
\providecommand \citenamefont [1]{#1}%
\providecommand \href@noop [0]{\@secondoftwo}%
\providecommand \href [0]{\begingroup \@sanitize@url \@href}%
\providecommand \@href[1]{\@@startlink{#1}\@@href}%
\providecommand \@@href[1]{\endgroup#1\@@endlink}%
\providecommand \@sanitize@url [0]{\catcode `\\12\catcode `\$12\catcode
  `\&12\catcode `\#12\catcode `\^12\catcode `\_12\catcode `\%12\relax}%
\providecommand \@@startlink[1]{}%
\providecommand \@@endlink[0]{}%
\providecommand \url  [0]{\begingroup\@sanitize@url \@url }%
\providecommand \@url [1]{\endgroup\@href {#1}{\urlprefix }}%
\providecommand \urlprefix  [0]{URL }%
\providecommand \Eprint [0]{\href }%
\providecommand \doibase [0]{http://dx.doi.org/}%
\providecommand \selectlanguage [0]{\@gobble}%
\providecommand \bibinfo  [0]{\@secondoftwo}%
\providecommand \bibfield  [0]{\@secondoftwo}%
\providecommand \translation [1]{[#1]}%
\providecommand \BibitemOpen [0]{}%
\providecommand \bibitemStop [0]{}%
\providecommand \bibitemNoStop [0]{.\EOS\space}%
\providecommand \EOS [0]{\spacefactor3000\relax}%
\providecommand \BibitemShut  [1]{\csname bibitem#1\endcsname}%
\let\auto@bib@innerbib\@empty
\bibitem [{\citenamefont {Barenco}\ \emph {et~al.}(1995)\citenamefont
  {Barenco}, \citenamefont {Bennett}, \citenamefont {Cleve}, \citenamefont
  {DiVincenzo}, \citenamefont {Margolus}, \citenamefont {Shor}, \citenamefont
  {Sleator}, \citenamefont {Smolin},\ and\ \citenamefont
  {Weinfurter}}]{Barenco1995}%
  \BibitemOpen
  \bibfield  {author} {\bibinfo {author} {\bibfnamefont {A.}~\bibnamefont
  {Barenco}}, \bibinfo {author} {\bibfnamefont {C.~H.}\ \bibnamefont
  {Bennett}}, \bibinfo {author} {\bibfnamefont {R.}~\bibnamefont {Cleve}},
  \bibinfo {author} {\bibfnamefont {D.~P.}\ \bibnamefont {DiVincenzo}},
  \bibinfo {author} {\bibfnamefont {N.}~\bibnamefont {Margolus}}, \bibinfo
  {author} {\bibfnamefont {P.}~\bibnamefont {Shor}}, \bibinfo {author}
  {\bibfnamefont {T.}~\bibnamefont {Sleator}}, \bibinfo {author} {\bibfnamefont
  {J.~A.}\ \bibnamefont {Smolin}}, \ and\ \bibinfo {author} {\bibfnamefont
  {H.}~\bibnamefont {Weinfurter}},\ }\bibfield  {title} {\enquote {\bibinfo
  {title} {Elementary gates for quantum computation},}\ }\href {\doibase
  10.1103/PhysRevA.52.3457} {\bibfield  {journal} {\bibinfo  {journal} {Phys.
  Rev. A}\ }\textbf {\bibinfo {volume} {52}},\ \bibinfo {pages} {3457--3467}
  (\bibinfo {year} {1995})}\BibitemShut {NoStop}%
\bibitem [{\citenamefont {Arute}\ \emph {et~al.}(2019)\citenamefont {Arute},
  \citenamefont {Arya}, \citenamefont {Babbush}, \citenamefont {Bacon},
  \citenamefont {Bardin}, \citenamefont {Barends}, \citenamefont {Biswas},
  \citenamefont {Boixo}, \citenamefont {Brandao}, \citenamefont {Buell},
  \citenamefont {Burkett}, \citenamefont {Chen}, \citenamefont {Chen},
  \citenamefont {Chiaro}, \citenamefont {Collins}, \citenamefont {Courtney},
  \citenamefont {Dunsworth}, \citenamefont {Farhi}, \citenamefont {Foxen},
  \citenamefont {Fowler}, \citenamefont {Gidney}, \citenamefont {Giustina},
  \citenamefont {Graff}, \citenamefont {Guerin}, \citenamefont {Habegger},
  \citenamefont {Harrigan}, \citenamefont {Hartmann}, \citenamefont {Ho},
  \citenamefont {Hoffmann}, \citenamefont {Huang}, \citenamefont {Humble},
  \citenamefont {Isakov}, \citenamefont {Jeffrey}, \citenamefont {Jiang},
  \citenamefont {Kafri}, \citenamefont {Kechedzhi}, \citenamefont {Kelly},
  \citenamefont {Klimov}, \citenamefont {Knysh}, \citenamefont {Korotkov},
  \citenamefont {Kostritsa}, \citenamefont {Landhuis}, \citenamefont
  {Lindmark}, \citenamefont {Lucero}, \citenamefont {Lyakh}, \citenamefont
  {Mandr\`a}, \citenamefont {McClean}, \citenamefont {McEwen}, \citenamefont
  {Megrant}, \citenamefont {Mi}, \citenamefont {Michielsen}, \citenamefont
  {Mohseni}, \citenamefont {Mutus}, \citenamefont {Naaman}, \citenamefont
  {Neeley}, \citenamefont {Neill}, \citenamefont {Niu}, \citenamefont {Ostby},
  \citenamefont {Petukhov}, \citenamefont {Platt}, \citenamefont {Quintana},
  \citenamefont {Rieffel}, \citenamefont {Roushan}, \citenamefont {Rubin},
  \citenamefont {Sank}, \citenamefont {Satzinger}, \citenamefont {Smelyanskiy},
  \citenamefont {Sung}, \citenamefont {Trevithick}, \citenamefont
  {Vainsencher}, \citenamefont {Villalonga}, \citenamefont {White},
  \citenamefont {Yao}, \citenamefont {Yeh}, \citenamefont {Zalcman},
  \citenamefont {Neven},\ and\ \citenamefont {Martinis}}]{Arute2019}%
  \BibitemOpen
  \bibfield  {author} {\bibinfo {author} {\bibfnamefont {F.}~\bibnamefont
  {Arute}}, \bibinfo {author} {\bibfnamefont {K.}~\bibnamefont {Arya}},
  \bibinfo {author} {\bibfnamefont {R.}~\bibnamefont {Babbush}}, \bibinfo
  {author} {\bibfnamefont {D.}~\bibnamefont {Bacon}}, \bibinfo {author}
  {\bibfnamefont {J.~C.}\ \bibnamefont {Bardin}}, \bibinfo {author}
  {\bibfnamefont {R.}~\bibnamefont {Barends}}, \bibinfo {author} {\bibfnamefont
  {R.}~\bibnamefont {Biswas}}, \bibinfo {author} {\bibfnamefont
  {S.}~\bibnamefont {Boixo}}, \bibinfo {author} {\bibfnamefont {F.~G. S.~L.}\
  \bibnamefont {Brandao}}, \bibinfo {author} {\bibfnamefont {D.~A.}\
  \bibnamefont {Buell}}, \bibinfo {author} {\bibfnamefont {B.}~\bibnamefont
  {Burkett}}, \bibinfo {author} {\bibfnamefont {Y.}~\bibnamefont {Chen}},
  \bibinfo {author} {\bibfnamefont {Z.}~\bibnamefont {Chen}}, \bibinfo {author}
  {\bibfnamefont {B.}~\bibnamefont {Chiaro}}, \bibinfo {author} {\bibfnamefont
  {R.}~\bibnamefont {Collins}}, \bibinfo {author} {\bibfnamefont
  {W.}~\bibnamefont {Courtney}}, \bibinfo {author} {\bibfnamefont
  {A.}~\bibnamefont {Dunsworth}}, \bibinfo {author} {\bibfnamefont
  {E.}~\bibnamefont {Farhi}}, \bibinfo {author} {\bibfnamefont
  {B.}~\bibnamefont {Foxen}}, \bibinfo {author} {\bibfnamefont
  {A.}~\bibnamefont {Fowler}}, \bibinfo {author} {\bibfnamefont
  {C.}~\bibnamefont {Gidney}}, \bibinfo {author} {\bibfnamefont
  {M.}~\bibnamefont {Giustina}}, \bibinfo {author} {\bibfnamefont
  {R.}~\bibnamefont {Graff}}, \bibinfo {author} {\bibfnamefont
  {K.}~\bibnamefont {Guerin}}, \bibinfo {author} {\bibfnamefont
  {S.}~\bibnamefont {Habegger}}, \bibinfo {author} {\bibfnamefont {M.~P.}\
  \bibnamefont {Harrigan}}, \bibinfo {author} {\bibfnamefont {M.~J.}\
  \bibnamefont {Hartmann}}, \bibinfo {author} {\bibfnamefont {A.}~\bibnamefont
  {Ho}}, \bibinfo {author} {\bibfnamefont {M.}~\bibnamefont {Hoffmann}},
  \bibinfo {author} {\bibfnamefont {T.}~\bibnamefont {Huang}}, \bibinfo
  {author} {\bibfnamefont {T.~S.}\ \bibnamefont {Humble}}, \bibinfo {author}
  {\bibfnamefont {S.~V.}\ \bibnamefont {Isakov}}, \bibinfo {author}
  {\bibfnamefont {E.}~\bibnamefont {Jeffrey}}, \bibinfo {author} {\bibfnamefont
  {Z.}~\bibnamefont {Jiang}}, \bibinfo {author} {\bibfnamefont
  {D.}~\bibnamefont {Kafri}}, \bibinfo {author} {\bibfnamefont
  {K.}~\bibnamefont {Kechedzhi}}, \bibinfo {author} {\bibfnamefont
  {J.}~\bibnamefont {Kelly}}, \bibinfo {author} {\bibfnamefont {P.~V.}\
  \bibnamefont {Klimov}}, \bibinfo {author} {\bibfnamefont {S.}~\bibnamefont
  {Knysh}}, \bibinfo {author} {\bibfnamefont {A.}~\bibnamefont {Korotkov}},
  \bibinfo {author} {\bibfnamefont {F.}~\bibnamefont {Kostritsa}}, \bibinfo
  {author} {\bibfnamefont {D.}~\bibnamefont {Landhuis}}, \bibinfo {author}
  {\bibfnamefont {M.}~\bibnamefont {Lindmark}}, \bibinfo {author}
  {\bibfnamefont {E.}~\bibnamefont {Lucero}}, \bibinfo {author} {\bibfnamefont
  {D.}~\bibnamefont {Lyakh}}, \bibinfo {author} {\bibfnamefont
  {S.}~\bibnamefont {Mandr\`a}}, \bibinfo {author} {\bibfnamefont {J.~R.}\
  \bibnamefont {McClean}}, \bibinfo {author} {\bibfnamefont {M.}~\bibnamefont
  {McEwen}}, \bibinfo {author} {\bibfnamefont {A.}~\bibnamefont {Megrant}},
  \bibinfo {author} {\bibfnamefont {X.}~\bibnamefont {Mi}}, \bibinfo {author}
  {\bibfnamefont {K.}~\bibnamefont {Michielsen}}, \bibinfo {author}
  {\bibfnamefont {M.}~\bibnamefont {Mohseni}}, \bibinfo {author} {\bibfnamefont
  {J.}~\bibnamefont {Mutus}}, \bibinfo {author} {\bibfnamefont
  {O.}~\bibnamefont {Naaman}}, \bibinfo {author} {\bibfnamefont
  {M.}~\bibnamefont {Neeley}}, \bibinfo {author} {\bibfnamefont
  {C.}~\bibnamefont {Neill}}, \bibinfo {author} {\bibfnamefont {M.~Y.}\
  \bibnamefont {Niu}}, \bibinfo {author} {\bibfnamefont {E.}~\bibnamefont
  {Ostby}}, \bibinfo {author} {\bibfnamefont {A.}~\bibnamefont {Petukhov}},
  \bibinfo {author} {\bibfnamefont {J.~C.}\ \bibnamefont {Platt}}, \bibinfo
  {author} {\bibfnamefont {C.}~\bibnamefont {Quintana}}, \bibinfo {author}
  {\bibfnamefont {E.~G.}\ \bibnamefont {Rieffel}}, \bibinfo {author}
  {\bibfnamefont {P.}~\bibnamefont {Roushan}}, \bibinfo {author} {\bibfnamefont
  {N.~C.}\ \bibnamefont {Rubin}}, \bibinfo {author} {\bibfnamefont
  {D.}~\bibnamefont {Sank}}, \bibinfo {author} {\bibfnamefont {K.~J.}\
  \bibnamefont {Satzinger}}, \bibinfo {author} {\bibfnamefont {V.}~\bibnamefont
  {Smelyanskiy}}, \bibinfo {author} {\bibfnamefont {K.~J.}\ \bibnamefont
  {Sung}}, \bibinfo {author} {\bibfnamefont {M.~D.}\ \bibnamefont
  {Trevithick}}, \bibinfo {author} {\bibfnamefont {A.}~\bibnamefont
  {Vainsencher}}, \bibinfo {author} {\bibfnamefont {B.}~\bibnamefont
  {Villalonga}}, \bibinfo {author} {\bibfnamefont {T.}~\bibnamefont {White}},
  \bibinfo {author} {\bibfnamefont {Z.~J.}\ \bibnamefont {Yao}}, \bibinfo
  {author} {\bibfnamefont {P.}~\bibnamefont {Yeh}}, \bibinfo {author}
  {\bibfnamefont {A.}~\bibnamefont {Zalcman}}, \bibinfo {author} {\bibfnamefont
  {H.}~\bibnamefont {Neven}}, \ and\ \bibinfo {author} {\bibfnamefont {J.~M.}\
  \bibnamefont {Martinis}},\ }\bibfield  {title} {\enquote {\bibinfo {title}
  {Quantum supremacy using a programmable superconducting processor},}\ }\href
  {\doibase 10.1038/s41586-019-1666-5} {\bibfield  {journal} {\bibinfo
  {journal} {Nature}\ }\textbf {\bibinfo {volume} {574}},\ \bibinfo {pages}
  {505--510} (\bibinfo {year} {2019})}\BibitemShut {NoStop}%
\bibitem [{\citenamefont {Jurcevic}\ \emph {et~al.}(2020)\citenamefont
  {Jurcevic}, \citenamefont {Javadi-Abhari}, \citenamefont {Bishop},
  \citenamefont {Lauer}, \citenamefont {Bogorin}, \citenamefont {Brink},
  \citenamefont {Capelluto}, \citenamefont {G\"unl\"uk}, \citenamefont {Itoko},
  \citenamefont {Kanazawa}, \citenamefont {Kandala}, \citenamefont {Keefe},
  \citenamefont {Krsulich}, \citenamefont {Landers}, \citenamefont
  {Lewandowski}, \citenamefont {McClure}, \citenamefont {Nannicini},
  \citenamefont {Narasgond}, \citenamefont {Nayfeh}, \citenamefont {Pritchett},
  \citenamefont {Rothwell}, \citenamefont {Srinivasan}, \citenamefont
  {Sundaresan}, \citenamefont {Wang}, \citenamefont {Wei}, \citenamefont
  {Wood}, \citenamefont {Yau}, \citenamefont {Zhang}, \citenamefont {Dial},
  \citenamefont {Chow},\ and\ \citenamefont {Gambetta}}]{Jurcevic2020}%
  \BibitemOpen
  \bibfield  {author} {\bibinfo {author} {\bibfnamefont {P.}~\bibnamefont
  {Jurcevic}}, \bibinfo {author} {\bibfnamefont {A.}~\bibnamefont
  {Javadi-Abhari}}, \bibinfo {author} {\bibfnamefont {L.~S.}\ \bibnamefont
  {Bishop}}, \bibinfo {author} {\bibfnamefont {I.}~\bibnamefont {Lauer}},
  \bibinfo {author} {\bibfnamefont {D.~F.}\ \bibnamefont {Bogorin}}, \bibinfo
  {author} {\bibfnamefont {M.}~\bibnamefont {Brink}}, \bibinfo {author}
  {\bibfnamefont {L.}~\bibnamefont {Capelluto}}, \bibinfo {author}
  {\bibfnamefont {O.}~\bibnamefont {G\"unl\"uk}}, \bibinfo {author}
  {\bibfnamefont {T.}~\bibnamefont {Itoko}}, \bibinfo {author} {\bibfnamefont
  {N.}~\bibnamefont {Kanazawa}}, \bibinfo {author} {\bibfnamefont
  {A.}~\bibnamefont {Kandala}}, \bibinfo {author} {\bibfnamefont {G.~A.}\
  \bibnamefont {Keefe}}, \bibinfo {author} {\bibfnamefont {K.}~\bibnamefont
  {Krsulich}}, \bibinfo {author} {\bibfnamefont {W.}~\bibnamefont {Landers}},
  \bibinfo {author} {\bibfnamefont {E.~P.}\ \bibnamefont {Lewandowski}},
  \bibinfo {author} {\bibfnamefont {D.~T.}\ \bibnamefont {McClure}}, \bibinfo
  {author} {\bibfnamefont {G.}~\bibnamefont {Nannicini}}, \bibinfo {author}
  {\bibfnamefont {A.}~\bibnamefont {Narasgond}}, \bibinfo {author}
  {\bibfnamefont {H.~M.}\ \bibnamefont {Nayfeh}}, \bibinfo {author}
  {\bibfnamefont {E.}~\bibnamefont {Pritchett}}, \bibinfo {author}
  {\bibfnamefont {M.~B.}\ \bibnamefont {Rothwell}}, \bibinfo {author}
  {\bibfnamefont {S.}~\bibnamefont {Srinivasan}}, \bibinfo {author}
  {\bibfnamefont {N.}~\bibnamefont {Sundaresan}}, \bibinfo {author}
  {\bibfnamefont {C.}~\bibnamefont {Wang}}, \bibinfo {author} {\bibfnamefont
  {K.~X.}\ \bibnamefont {Wei}}, \bibinfo {author} {\bibfnamefont {C.~J.}\
  \bibnamefont {Wood}}, \bibinfo {author} {\bibfnamefont {J.-B.}\ \bibnamefont
  {Yau}}, \bibinfo {author} {\bibfnamefont {E.~J.}\ \bibnamefont {Zhang}},
  \bibinfo {author} {\bibfnamefont {O.~E.}\ \bibnamefont {Dial}}, \bibinfo
  {author} {\bibfnamefont {J.~M.}\ \bibnamefont {Chow}}, \ and\ \bibinfo
  {author} {\bibfnamefont {J.~M.}\ \bibnamefont {Gambetta}},\ }\href@noop {}
  {\enquote {\bibinfo {title} {Demonstration of quantum volume 64 on a
  superconducting quantum computing system},}\ } (\bibinfo {year} {2020}),\
  \Eprint {http://arxiv.org/abs/2008.08571} {arXiv:2008.08571 [quant-ph]}
  \BibitemShut {NoStop}%
\bibitem [{\citenamefont {Wu}\ \emph {et~al.}(2021)\citenamefont {Wu},
  \citenamefont {Bao}, \citenamefont {Cao}, \citenamefont {Chen}, \citenamefont
  {Chen}, \citenamefont {Chen}, \citenamefont {Chung}, \citenamefont {Deng},
  \citenamefont {Du}, \citenamefont {Fan}, \citenamefont {Gong}, \citenamefont
  {Guo}, \citenamefont {Guo}, \citenamefont {Guo}, \citenamefont {Han},
  \citenamefont {Hong}, \citenamefont {Huang}, \citenamefont {Huo},
  \citenamefont {Li}, \citenamefont {Li}, \citenamefont {Li}, \citenamefont
  {Li}, \citenamefont {Liang}, \citenamefont {Lin}, \citenamefont {Lin},
  \citenamefont {Qian}, \citenamefont {Qiao}, \citenamefont {Rong},
  \citenamefont {Su}, \citenamefont {Sun}, \citenamefont {Wang}, \citenamefont
  {Wang}, \citenamefont {Wu}, \citenamefont {Xu}, \citenamefont {Yan},
  \citenamefont {Yang}, \citenamefont {Yang}, \citenamefont {Ye}, \citenamefont
  {Yin}, \citenamefont {Ying}, \citenamefont {Yu}, \citenamefont {Zha},
  \citenamefont {Zhang}, \citenamefont {Zhang}, \citenamefont {Zhang},
  \citenamefont {Zhang}, \citenamefont {Zhao}, \citenamefont {Zhao},
  \citenamefont {Zhou}, \citenamefont {Zhu}, \citenamefont {Lu}, \citenamefont
  {Peng}, \citenamefont {Zhu},\ and\ \citenamefont {Pan}}]{Zhu2021}%
  \BibitemOpen
  \bibfield  {author} {\bibinfo {author} {\bibfnamefont {Y.}~\bibnamefont
  {Wu}}, \bibinfo {author} {\bibfnamefont {W.-S.}\ \bibnamefont {Bao}},
  \bibinfo {author} {\bibfnamefont {S.}~\bibnamefont {Cao}}, \bibinfo {author}
  {\bibfnamefont {F.}~\bibnamefont {Chen}}, \bibinfo {author} {\bibfnamefont
  {M.-C.}\ \bibnamefont {Chen}}, \bibinfo {author} {\bibfnamefont
  {X.}~\bibnamefont {Chen}}, \bibinfo {author} {\bibfnamefont {T.-H.}\
  \bibnamefont {Chung}}, \bibinfo {author} {\bibfnamefont {H.}~\bibnamefont
  {Deng}}, \bibinfo {author} {\bibfnamefont {Y.}~\bibnamefont {Du}}, \bibinfo
  {author} {\bibfnamefont {D.}~\bibnamefont {Fan}}, \bibinfo {author}
  {\bibfnamefont {M.}~\bibnamefont {Gong}}, \bibinfo {author} {\bibfnamefont
  {C.}~\bibnamefont {Guo}}, \bibinfo {author} {\bibfnamefont {C.}~\bibnamefont
  {Guo}}, \bibinfo {author} {\bibfnamefont {S.}~\bibnamefont {Guo}}, \bibinfo
  {author} {\bibfnamefont {L.}~\bibnamefont {Han}}, \bibinfo {author}
  {\bibfnamefont {L.}~\bibnamefont {Hong}}, \bibinfo {author} {\bibfnamefont
  {H.-L.}\ \bibnamefont {Huang}}, \bibinfo {author} {\bibfnamefont {Y.-H.}\
  \bibnamefont {Huo}}, \bibinfo {author} {\bibfnamefont {L.}~\bibnamefont
  {Li}}, \bibinfo {author} {\bibfnamefont {N.}~\bibnamefont {Li}}, \bibinfo
  {author} {\bibfnamefont {S.}~\bibnamefont {Li}}, \bibinfo {author}
  {\bibfnamefont {Y.}~\bibnamefont {Li}}, \bibinfo {author} {\bibfnamefont
  {F.}~\bibnamefont {Liang}}, \bibinfo {author} {\bibfnamefont
  {C.}~\bibnamefont {Lin}}, \bibinfo {author} {\bibfnamefont {J.}~\bibnamefont
  {Lin}}, \bibinfo {author} {\bibfnamefont {H.}~\bibnamefont {Qian}}, \bibinfo
  {author} {\bibfnamefont {D.}~\bibnamefont {Qiao}}, \bibinfo {author}
  {\bibfnamefont {H.}~\bibnamefont {Rong}}, \bibinfo {author} {\bibfnamefont
  {H.}~\bibnamefont {Su}}, \bibinfo {author} {\bibfnamefont {L.}~\bibnamefont
  {Sun}}, \bibinfo {author} {\bibfnamefont {L.}~\bibnamefont {Wang}}, \bibinfo
  {author} {\bibfnamefont {S.}~\bibnamefont {Wang}}, \bibinfo {author}
  {\bibfnamefont {D.}~\bibnamefont {Wu}}, \bibinfo {author} {\bibfnamefont
  {Y.}~\bibnamefont {Xu}}, \bibinfo {author} {\bibfnamefont {K.}~\bibnamefont
  {Yan}}, \bibinfo {author} {\bibfnamefont {W.}~\bibnamefont {Yang}}, \bibinfo
  {author} {\bibfnamefont {Y.}~\bibnamefont {Yang}}, \bibinfo {author}
  {\bibfnamefont {Y.}~\bibnamefont {Ye}}, \bibinfo {author} {\bibfnamefont
  {J.}~\bibnamefont {Yin}}, \bibinfo {author} {\bibfnamefont {C.}~\bibnamefont
  {Ying}}, \bibinfo {author} {\bibfnamefont {J.}~\bibnamefont {Yu}}, \bibinfo
  {author} {\bibfnamefont {C.}~\bibnamefont {Zha}}, \bibinfo {author}
  {\bibfnamefont {C.}~\bibnamefont {Zhang}}, \bibinfo {author} {\bibfnamefont
  {H.}~\bibnamefont {Zhang}}, \bibinfo {author} {\bibfnamefont
  {K.}~\bibnamefont {Zhang}}, \bibinfo {author} {\bibfnamefont
  {Y.}~\bibnamefont {Zhang}}, \bibinfo {author} {\bibfnamefont
  {H.}~\bibnamefont {Zhao}}, \bibinfo {author} {\bibfnamefont {Y.}~\bibnamefont
  {Zhao}}, \bibinfo {author} {\bibfnamefont {L.}~\bibnamefont {Zhou}}, \bibinfo
  {author} {\bibfnamefont {Q.}~\bibnamefont {Zhu}}, \bibinfo {author}
  {\bibfnamefont {C.-Y.}\ \bibnamefont {Lu}}, \bibinfo {author} {\bibfnamefont
  {C.-Z.}\ \bibnamefont {Peng}}, \bibinfo {author} {\bibfnamefont
  {X.}~\bibnamefont {Zhu}}, \ and\ \bibinfo {author} {\bibfnamefont {J.-W.}\
  \bibnamefont {Pan}},\ }\href@noop {} {\enquote {\bibinfo {title} {Strong
  quantum computational advantage using a superconducting quantum processor},}\
  } (\bibinfo {year} {2021}),\ \Eprint {http://arxiv.org/abs/2106.14734}
  {arXiv:2106.14734 [quant-ph]} \BibitemShut {NoStop}%
\bibitem [{\citenamefont {Babbush}\ \emph {et~al.}(2018)\citenamefont
  {Babbush}, \citenamefont {Wiebe}, \citenamefont {McClean}, \citenamefont
  {McClain}, \citenamefont {Neven},\ and\ \citenamefont {Chan}}]{Babbush2018}%
  \BibitemOpen
  \bibfield  {author} {\bibinfo {author} {\bibfnamefont {R.}~\bibnamefont
  {Babbush}}, \bibinfo {author} {\bibfnamefont {N.}~\bibnamefont {Wiebe}},
  \bibinfo {author} {\bibfnamefont {J.}~\bibnamefont {McClean}}, \bibinfo
  {author} {\bibfnamefont {J.}~\bibnamefont {McClain}}, \bibinfo {author}
  {\bibfnamefont {H.}~\bibnamefont {Neven}}, \ and\ \bibinfo {author}
  {\bibfnamefont {G.~K.-L.}\ \bibnamefont {Chan}},\ }\bibfield  {title}
  {\enquote {\bibinfo {title} {Low-depth quantum simulation of materials},}\
  }\href {\doibase 10.1103/PhysRevX.8.011044} {\bibfield  {journal} {\bibinfo
  {journal} {Phys. Rev. X}\ }\textbf {\bibinfo {volume} {8}},\ \bibinfo {pages}
  {011044} (\bibinfo {year} {2018})}\BibitemShut {NoStop}%
\bibitem [{\citenamefont {Bharti}\ \emph {et~al.}(2021)\citenamefont {Bharti},
  \citenamefont {Cervera-Lierta}, \citenamefont {Kyaw}, \citenamefont {Haug},
  \citenamefont {Alperin-Lea}, \citenamefont {Anand}, \citenamefont {Degroote},
  \citenamefont {Heimonen}, \citenamefont {Kottmann}, \citenamefont {Menke},
  \citenamefont {Mok}, \citenamefont {Sim}, \citenamefont {Kwek},\ and\
  \citenamefont {Aspuru-Guzik}}]{Bharti2021}%
  \BibitemOpen
  \bibfield  {author} {\bibinfo {author} {\bibfnamefont {K.}~\bibnamefont
  {Bharti}}, \bibinfo {author} {\bibfnamefont {A.}~\bibnamefont
  {Cervera-Lierta}}, \bibinfo {author} {\bibfnamefont {T.~H.}\ \bibnamefont
  {Kyaw}}, \bibinfo {author} {\bibfnamefont {T.}~\bibnamefont {Haug}}, \bibinfo
  {author} {\bibfnamefont {S.}~\bibnamefont {Alperin-Lea}}, \bibinfo {author}
  {\bibfnamefont {A.}~\bibnamefont {Anand}}, \bibinfo {author} {\bibfnamefont
  {M.}~\bibnamefont {Degroote}}, \bibinfo {author} {\bibfnamefont
  {H.}~\bibnamefont {Heimonen}}, \bibinfo {author} {\bibfnamefont {J.~S.}\
  \bibnamefont {Kottmann}}, \bibinfo {author} {\bibfnamefont {T.}~\bibnamefont
  {Menke}}, \bibinfo {author} {\bibfnamefont {W.-K.}\ \bibnamefont {Mok}},
  \bibinfo {author} {\bibfnamefont {S.}~\bibnamefont {Sim}}, \bibinfo {author}
  {\bibfnamefont {L.-C.}\ \bibnamefont {Kwek}}, \ and\ \bibinfo {author}
  {\bibfnamefont {A.}~\bibnamefont {Aspuru-Guzik}},\ }\bibfield  {title}
  {\enquote {\bibinfo {title} {Noisy intermediate-scale quantum (nisq)
  algorithms},}\ }\href@noop {} {\  (\bibinfo {year} {2021})},\ \Eprint
  {http://arxiv.org/abs/arXiv:2101.08448v1} {arXiv:2101.08448v1} \BibitemShut
  {NoStop}%
\bibitem [{\citenamefont {Peng}\ \emph {et~al.}(2009)\citenamefont {Peng},
  \citenamefont {Zhang}, \citenamefont {Du},\ and\ \citenamefont
  {Suter}}]{Peng2009}%
  \BibitemOpen
  \bibfield  {author} {\bibinfo {author} {\bibfnamefont {X.}~\bibnamefont
  {Peng}}, \bibinfo {author} {\bibfnamefont {J.}~\bibnamefont {Zhang}},
  \bibinfo {author} {\bibfnamefont {J.}~\bibnamefont {Du}}, \ and\ \bibinfo
  {author} {\bibfnamefont {D.}~\bibnamefont {Suter}},\ }\bibfield  {title}
  {\enquote {\bibinfo {title} {Quantum simulation of a system with competing
  two- and three-body interactions},}\ }\href {\doibase
  10.1103/PhysRevLett.103.140501} {\bibfield  {journal} {\bibinfo  {journal}
  {Phys. Rev. Lett.}\ }\textbf {\bibinfo {volume} {103}},\ \bibinfo {pages}
  {140501} (\bibinfo {year} {2009})}\BibitemShut {NoStop}%
\bibitem [{\citenamefont {Georgescu}\ \emph {et~al.}(2014)\citenamefont
  {Georgescu}, \citenamefont {Ashhab},\ and\ \citenamefont
  {Nori}}]{Georgescu2014}%
  \BibitemOpen
  \bibfield  {author} {\bibinfo {author} {\bibfnamefont {I.~M.}\ \bibnamefont
  {Georgescu}}, \bibinfo {author} {\bibfnamefont {S.}~\bibnamefont {Ashhab}}, \
  and\ \bibinfo {author} {\bibfnamefont {F.}~\bibnamefont {Nori}},\ }\bibfield
  {title} {\enquote {\bibinfo {title} {Quantum simulation},}\ }\href {\doibase
  10.1103/RevModPhys.86.153} {\bibfield  {journal} {\bibinfo  {journal} {Rev.
  Mod. Phys.}\ }\textbf {\bibinfo {volume} {86}},\ \bibinfo {pages} {153--185}
  (\bibinfo {year} {2014})}\BibitemShut {NoStop}%
\bibitem [{\citenamefont {Dzialoshinskii}(1957)}]{Dzy1957}%
  \BibitemOpen
  \bibfield  {author} {\bibinfo {author} {\bibfnamefont {I.~E.}\ \bibnamefont
  {Dzialoshinskii}},\ }\bibfield  {title} {\enquote {\bibinfo {title}
  {{Thermodynamic theory of weak ferromagnetism in antiferromagnetic
  substances}},}\ }\href@noop {} {\bibfield  {journal} {\bibinfo  {journal} {J.
  Exp. Theoret. Phys.}\ }\textbf {\bibinfo {volume} {32}},\ \bibinfo {pages}
  {1547--1562} (\bibinfo {year} {1957})}\BibitemShut {NoStop}%
\bibitem [{\citenamefont {{Moriya}}(1960)}]{Moriya1960}%
  \BibitemOpen
  \bibfield  {author} {\bibinfo {author} {\bibfnamefont {T.}~\bibnamefont
  {{Moriya}}},\ }\bibfield  {title} {\enquote {\bibinfo {title} {{New Mechanism
  of Anisotropic Superexchange Interaction}},}\ }\href {\doibase
  10.1103/PhysRevLett.4.228} {\bibfield  {journal} {\bibinfo  {journal} {Phys.
  Rev. Lett.}\ }\textbf {\bibinfo {volume} {4}},\ \bibinfo {pages} {228--230}
  (\bibinfo {year} {1960})}\BibitemShut {NoStop}%
\bibitem [{\citenamefont {Wang}\ \emph {et~al.}(2019)\citenamefont {Wang},
  \citenamefont {Song}, \citenamefont {Feng}, \citenamefont {Cai},
  \citenamefont {Xu}, \citenamefont {Deng}, \citenamefont {Li}, \citenamefont
  {Zheng}, \citenamefont {Zhu}, \citenamefont {Wang}, \citenamefont {Zhu},\
  and\ \citenamefont {Scully}}]{Wang2019}%
  \BibitemOpen
  \bibfield  {author} {\bibinfo {author} {\bibfnamefont {D.-W.}\ \bibnamefont
  {Wang}}, \bibinfo {author} {\bibfnamefont {C.}~\bibnamefont {Song}}, \bibinfo
  {author} {\bibfnamefont {W.}~\bibnamefont {Feng}}, \bibinfo {author}
  {\bibfnamefont {H.}~\bibnamefont {Cai}}, \bibinfo {author} {\bibfnamefont
  {D.}~\bibnamefont {Xu}}, \bibinfo {author} {\bibfnamefont {H.}~\bibnamefont
  {Deng}}, \bibinfo {author} {\bibfnamefont {H.}~\bibnamefont {Li}}, \bibinfo
  {author} {\bibfnamefont {D.}~\bibnamefont {Zheng}}, \bibinfo {author}
  {\bibfnamefont {X.}~\bibnamefont {Zhu}}, \bibinfo {author} {\bibfnamefont
  {H.}~\bibnamefont {Wang}}, \bibinfo {author} {\bibfnamefont {S.-Y.}\
  \bibnamefont {Zhu}}, \ and\ \bibinfo {author} {\bibfnamefont {M.~O.}\
  \bibnamefont {Scully}},\ }\bibfield  {title} {\enquote {\bibinfo {title}
  {Synthesis of antisymmetric spin exchange interaction and chiral spin
  clusters in superconducting circuits},}\ }\href {\doibase
  10.1038/s41567-018-0400-9} {\bibfield  {journal} {\bibinfo  {journal} {Nature
  Physics}\ }\textbf {\bibinfo {volume} {15}},\ \bibinfo {pages} {382--386}
  (\bibinfo {year} {2019})}\BibitemShut {NoStop}%
\bibitem [{\citenamefont {Kelly}\ \emph {et~al.}(2015)\citenamefont {Kelly},
  \citenamefont {Barends}, \citenamefont {Fowler}, \citenamefont {Megrant},
  \citenamefont {Jeffrey}, \citenamefont {White}, \citenamefont {Sank},
  \citenamefont {Mutus}, \citenamefont {Campbell}, \citenamefont {Chen},
  \citenamefont {Chen}, \citenamefont {Chiaro}, \citenamefont {Dunsworth},
  \citenamefont {Hoi}, \citenamefont {Neill}, \citenamefont {O'Malley},
  \citenamefont {Quintana}, \citenamefont {Roushan}, \citenamefont
  {Vainsencher}, \citenamefont {Wenner}, \citenamefont {Cleland},\ and\
  \citenamefont {Martinis}}]{Kelly2015}%
  \BibitemOpen
  \bibfield  {author} {\bibinfo {author} {\bibfnamefont {J.}~\bibnamefont
  {Kelly}}, \bibinfo {author} {\bibfnamefont {R.}~\bibnamefont {Barends}},
  \bibinfo {author} {\bibfnamefont {A.~G.}\ \bibnamefont {Fowler}}, \bibinfo
  {author} {\bibfnamefont {A.}~\bibnamefont {Megrant}}, \bibinfo {author}
  {\bibfnamefont {E.}~\bibnamefont {Jeffrey}}, \bibinfo {author} {\bibfnamefont
  {T.~C.}\ \bibnamefont {White}}, \bibinfo {author} {\bibfnamefont
  {D.}~\bibnamefont {Sank}}, \bibinfo {author} {\bibfnamefont {J.~Y.}\
  \bibnamefont {Mutus}}, \bibinfo {author} {\bibfnamefont {B.}~\bibnamefont
  {Campbell}}, \bibinfo {author} {\bibfnamefont {Y.}~\bibnamefont {Chen}},
  \bibinfo {author} {\bibfnamefont {Z.}~\bibnamefont {Chen}}, \bibinfo {author}
  {\bibfnamefont {B.}~\bibnamefont {Chiaro}}, \bibinfo {author} {\bibfnamefont
  {A.}~\bibnamefont {Dunsworth}}, \bibinfo {author} {\bibfnamefont {I.-C.}\
  \bibnamefont {Hoi}}, \bibinfo {author} {\bibfnamefont {C.}~\bibnamefont
  {Neill}}, \bibinfo {author} {\bibfnamefont {P.~J.~J.}\ \bibnamefont
  {O'Malley}}, \bibinfo {author} {\bibfnamefont {C.}~\bibnamefont {Quintana}},
  \bibinfo {author} {\bibfnamefont {P.}~\bibnamefont {Roushan}}, \bibinfo
  {author} {\bibfnamefont {A.}~\bibnamefont {Vainsencher}}, \bibinfo {author}
  {\bibfnamefont {J.}~\bibnamefont {Wenner}}, \bibinfo {author} {\bibfnamefont
  {A.~N.}\ \bibnamefont {Cleland}}, \ and\ \bibinfo {author} {\bibfnamefont
  {J.~M.}\ \bibnamefont {Martinis}},\ }\bibfield  {title} {\enquote {\bibinfo
  {title} {State preservation by repetitive error detection in a
  superconducting quantum circuit},}\ }\href {\doibase 10.1038/nature14270}
  {\bibfield  {journal} {\bibinfo  {journal} {Nature}\ }\textbf {\bibinfo
  {volume} {519}},\ \bibinfo {pages} {66--69} (\bibinfo {year}
  {2015})}\BibitemShut {NoStop}%
\bibitem [{\citenamefont {Roushan}\ \emph {et~al.}(2017)\citenamefont
  {Roushan}, \citenamefont {Neill}, \citenamefont {Megrant}, \citenamefont
  {Chen}, \citenamefont {Babbush}, \citenamefont {Barends}, \citenamefont
  {Campbell}, \citenamefont {Chen}, \citenamefont {Chiaro}, \citenamefont
  {Dunsworth}, \citenamefont {Fowler}, \citenamefont {Jeffrey}, \citenamefont
  {Kelly}, \citenamefont {Lucero}, \citenamefont {Mutus}, \citenamefont
  {O'Malley}, \citenamefont {Neeley}, \citenamefont {Quintana}, \citenamefont
  {Sank}, \citenamefont {Vainsencher}, \citenamefont {Wenner}, \citenamefont
  {White}, \citenamefont {Kapit}, \citenamefont {Neven},\ and\ \citenamefont
  {Martinis}}]{Roushan2017}%
  \BibitemOpen
  \bibfield  {author} {\bibinfo {author} {\bibfnamefont {P.}~\bibnamefont
  {Roushan}}, \bibinfo {author} {\bibfnamefont {C.}~\bibnamefont {Neill}},
  \bibinfo {author} {\bibfnamefont {A.}~\bibnamefont {Megrant}}, \bibinfo
  {author} {\bibfnamefont {Y.}~\bibnamefont {Chen}}, \bibinfo {author}
  {\bibfnamefont {R.}~\bibnamefont {Babbush}}, \bibinfo {author} {\bibfnamefont
  {R.}~\bibnamefont {Barends}}, \bibinfo {author} {\bibfnamefont
  {B.}~\bibnamefont {Campbell}}, \bibinfo {author} {\bibfnamefont
  {Z.}~\bibnamefont {Chen}}, \bibinfo {author} {\bibfnamefont {B.}~\bibnamefont
  {Chiaro}}, \bibinfo {author} {\bibfnamefont {A.}~\bibnamefont {Dunsworth}},
  \bibinfo {author} {\bibfnamefont {A.}~\bibnamefont {Fowler}}, \bibinfo
  {author} {\bibfnamefont {E.}~\bibnamefont {Jeffrey}}, \bibinfo {author}
  {\bibfnamefont {J.}~\bibnamefont {Kelly}}, \bibinfo {author} {\bibfnamefont
  {E.}~\bibnamefont {Lucero}}, \bibinfo {author} {\bibfnamefont
  {J.}~\bibnamefont {Mutus}}, \bibinfo {author} {\bibfnamefont {P.~J.~J.}\
  \bibnamefont {O'Malley}}, \bibinfo {author} {\bibfnamefont {M.}~\bibnamefont
  {Neeley}}, \bibinfo {author} {\bibfnamefont {C.}~\bibnamefont {Quintana}},
  \bibinfo {author} {\bibfnamefont {D.}~\bibnamefont {Sank}}, \bibinfo {author}
  {\bibfnamefont {A.}~\bibnamefont {Vainsencher}}, \bibinfo {author}
  {\bibfnamefont {J.}~\bibnamefont {Wenner}}, \bibinfo {author} {\bibfnamefont
  {T.}~\bibnamefont {White}}, \bibinfo {author} {\bibfnamefont
  {E.}~\bibnamefont {Kapit}}, \bibinfo {author} {\bibfnamefont
  {H.}~\bibnamefont {Neven}}, \ and\ \bibinfo {author} {\bibfnamefont
  {J.}~\bibnamefont {Martinis}},\ }\bibfield  {title} {\enquote {\bibinfo
  {title} {Chiral ground-state currents of interacting photons in a synthetic
  magnetic field},}\ }\href {\doibase 10.1038/nphys3930} {\bibfield  {journal}
  {\bibinfo  {journal} {Nature Physics}\ }\textbf {\bibinfo {volume} {13}},\
  \bibinfo {pages} {146--151} (\bibinfo {year} {2017})}\BibitemShut {NoStop}%
\bibitem [{\citenamefont {Liu}\ \emph {et~al.}(2020)\citenamefont {Liu},
  \citenamefont {Feng}, \citenamefont {Ren}, \citenamefont {Wang},\ and\
  \citenamefont {Wang}}]{Liu2020}%
  \BibitemOpen
  \bibfield  {author} {\bibinfo {author} {\bibfnamefont {W.}~\bibnamefont
  {Liu}}, \bibinfo {author} {\bibfnamefont {W.}~\bibnamefont {Feng}}, \bibinfo
  {author} {\bibfnamefont {W.}~\bibnamefont {Ren}}, \bibinfo {author}
  {\bibfnamefont {D.-W.}\ \bibnamefont {Wang}}, \ and\ \bibinfo {author}
  {\bibfnamefont {H.}~\bibnamefont {Wang}},\ }\bibfield  {title} {\enquote
  {\bibinfo {title} {Synthesizing three-body interaction of spin chirality with
  superconducting qubits},}\ }\href {\doibase 10.1063/1.5140884} {\bibfield
  {journal} {\bibinfo  {journal} {Applied Physics Letters}\ }\textbf {\bibinfo
  {volume} {116}},\ \bibinfo {pages} {114001} (\bibinfo {year}
  {2020})}\BibitemShut {NoStop}%
\bibitem [{\citenamefont {Wen}\ \emph {et~al.}(1989)\citenamefont {Wen},
  \citenamefont {Wilczek},\ and\ \citenamefont {Zee}}]{Wen1989}%
  \BibitemOpen
  \bibfield  {author} {\bibinfo {author} {\bibfnamefont {X.~G.}\ \bibnamefont
  {Wen}}, \bibinfo {author} {\bibfnamefont {F.}~\bibnamefont {Wilczek}}, \ and\
  \bibinfo {author} {\bibfnamefont {A.}~\bibnamefont {Zee}},\ }\bibfield
  {title} {\enquote {\bibinfo {title} {Chiral spin states and
  superconductivity},}\ }\href {\doibase 10.1103/PhysRevB.39.11413} {\bibfield
  {journal} {\bibinfo  {journal} {Phys. Rev. B}\ }\textbf {\bibinfo {volume}
  {39}},\ \bibinfo {pages} {11413--11423} (\bibinfo {year} {1989})}\BibitemShut
  {NoStop}%
\bibitem [{\citenamefont {Cai}\ and\ \citenamefont {Wang}(2021)}]{Cai2021}%
  \BibitemOpen
  \bibfield  {author} {\bibinfo {author} {\bibfnamefont {H.}~\bibnamefont
  {Cai}}\ and\ \bibinfo {author} {\bibfnamefont {D.-W.}\ \bibnamefont {Wang}},\
  }\bibfield  {title} {\enquote {\bibinfo {title} {Topological phases of
  quantized light},}\ }\href {\doibase 10.1093/nsr/nwaa196} {\bibfield
  {journal} {\bibinfo  {journal} {National Science Review}\ }\textbf {\bibinfo
  {volume} {8}},\ \bibinfo {pages} {nwaa196} (\bibinfo {year}
  {2021})}\BibitemShut {NoStop}%
\bibitem [{\citenamefont {Yuan}\ \emph {et~al.}(2021)\citenamefont {Yuan},
  \citenamefont {Cai}, \citenamefont {Wu}, \citenamefont {Zhu}, \citenamefont
  {Liu},\ and\ \citenamefont {Wang}}]{Yuan2021}%
  \BibitemOpen
  \bibfield  {author} {\bibinfo {author} {\bibfnamefont {J.}~\bibnamefont
  {Yuan}}, \bibinfo {author} {\bibfnamefont {H.}~\bibnamefont {Cai}}, \bibinfo
  {author} {\bibfnamefont {C.}~\bibnamefont {Wu}}, \bibinfo {author}
  {\bibfnamefont {S.-Y.}\ \bibnamefont {Zhu}}, \bibinfo {author} {\bibfnamefont
  {R.-B.}\ \bibnamefont {Liu}}, \ and\ \bibinfo {author} {\bibfnamefont
  {D.-W.}\ \bibnamefont {Wang}},\ }\bibfield  {title} {\enquote {\bibinfo
  {title} {Unification of valley and anomalous hall effects in a strained
  lattice},}\ }\href {\doibase 10.1103/PhysRevB.104.035410} {\bibfield
  {journal} {\bibinfo  {journal} {Phys. Rev. B}\ }\textbf {\bibinfo {volume}
  {104}},\ \bibinfo {pages} {035410} (\bibinfo {year} {2021})}\BibitemShut
  {NoStop}%
\bibitem [{\citenamefont {Cheng}\ \emph {et~al.}(2021)\citenamefont {Cheng},
  \citenamefont {Peng}, \citenamefont {Wang}, \citenamefont {Chen},
  \citenamefont {Yuan},\ and\ \citenamefont {Fan}}]{Cheng2021}%
  \BibitemOpen
  \bibfield  {author} {\bibinfo {author} {\bibfnamefont {D.}~\bibnamefont
  {Cheng}}, \bibinfo {author} {\bibfnamefont {B.}~\bibnamefont {Peng}},
  \bibinfo {author} {\bibfnamefont {D.-W.}\ \bibnamefont {Wang}}, \bibinfo
  {author} {\bibfnamefont {X.}~\bibnamefont {Chen}}, \bibinfo {author}
  {\bibfnamefont {L.}~\bibnamefont {Yuan}}, \ and\ \bibinfo {author}
  {\bibfnamefont {S.}~\bibnamefont {Fan}},\ }\bibfield  {title} {\enquote
  {\bibinfo {title} {Arbitrary synthetic dimensions via multiboson dynamics on
  a one-dimensional lattice},}\ }\href {\doibase
  10.1103/PhysRevResearch.3.033069} {\bibfield  {journal} {\bibinfo  {journal}
  {Phys. Rev. Research}\ }\textbf {\bibinfo {volume} {3}},\ \bibinfo {pages}
  {033069} (\bibinfo {year} {2021})}\BibitemShut {NoStop}%
\bibitem [{\citenamefont {Dai}\ \emph {et~al.}(2017)\citenamefont {Dai},
  \citenamefont {Yang}, \citenamefont {Reingruber}, \citenamefont {Sun},
  \citenamefont {Xu}, \citenamefont {Chen}, \citenamefont {Yuan},\ and\
  \citenamefont {Pan}}]{Dai2017}%
  \BibitemOpen
  \bibfield  {author} {\bibinfo {author} {\bibfnamefont {H.-N.}\ \bibnamefont
  {Dai}}, \bibinfo {author} {\bibfnamefont {B.}~\bibnamefont {Yang}}, \bibinfo
  {author} {\bibfnamefont {A.}~\bibnamefont {Reingruber}}, \bibinfo {author}
  {\bibfnamefont {H.}~\bibnamefont {Sun}}, \bibinfo {author} {\bibfnamefont
  {X.-F.}\ \bibnamefont {Xu}}, \bibinfo {author} {\bibfnamefont {Y.-A.}\
  \bibnamefont {Chen}}, \bibinfo {author} {\bibfnamefont {Z.-S.}\ \bibnamefont
  {Yuan}}, \ and\ \bibinfo {author} {\bibfnamefont {J.-W.}\ \bibnamefont
  {Pan}},\ }\bibfield  {title} {\enquote {\bibinfo {title} {Four-body
  ring-exchange interactions and anyonic statistics within a minimal toric-code
  hamiltonian},}\ }\href {\doibase 10.1038/nphys4243} {\bibfield  {journal}
  {\bibinfo  {journal} {Nature Physics}\ }\textbf {\bibinfo {volume} {13}},\
  \bibinfo {pages} {1195--1200} (\bibinfo {year} {2017})}\BibitemShut {NoStop}%
\bibitem [{\citenamefont {Kitaev}(2003)}]{Kitaev2003}%
  \BibitemOpen
  \bibfield  {author} {\bibinfo {author} {\bibfnamefont {A.}~\bibnamefont
  {Kitaev}},\ }\bibfield  {title} {\enquote {\bibinfo {title} {Fault-tolerant
  quantum computation by anyons},}\ }\href {\doibase
  10.1016/s0003-4916(02)00018-0} {\bibfield  {journal} {\bibinfo  {journal}
  {Annals of Physics}\ }\textbf {\bibinfo {volume} {303}},\ \bibinfo {pages}
  {2--30} (\bibinfo {year} {2003})}\BibitemShut {NoStop}%
\bibitem [{\citenamefont {Garziano}\ \emph {et~al.}(2016)\citenamefont
  {Garziano}, \citenamefont {Macr\`{\i}}, \citenamefont {Stassi}, \citenamefont
  {Di~Stefano}, \citenamefont {Nori},\ and\ \citenamefont
  {Savasta}}]{Garziano2016}%
  \BibitemOpen
  \bibfield  {author} {\bibinfo {author} {\bibfnamefont {L.}~\bibnamefont
  {Garziano}}, \bibinfo {author} {\bibfnamefont {V.}~\bibnamefont
  {Macr\`{\i}}}, \bibinfo {author} {\bibfnamefont {R.}~\bibnamefont {Stassi}},
  \bibinfo {author} {\bibfnamefont {O.}~\bibnamefont {Di~Stefano}}, \bibinfo
  {author} {\bibfnamefont {F.}~\bibnamefont {Nori}}, \ and\ \bibinfo {author}
  {\bibfnamefont {S.}~\bibnamefont {Savasta}},\ }\bibfield  {title} {\enquote
  {\bibinfo {title} {One photon can simultaneously excite two or more atoms},}\
  }\href {\doibase 10.1103/PhysRevLett.117.043601} {\bibfield  {journal}
  {\bibinfo  {journal} {Phys. Rev. Lett.}\ }\textbf {\bibinfo {volume} {117}},\
  \bibinfo {pages} {043601} (\bibinfo {year} {2016})}\BibitemShut {NoStop}%
\bibitem [{\citenamefont {Kockum}\ \emph {et~al.}(2017)\citenamefont {Kockum},
  \citenamefont {Miranowicz}, \citenamefont {Macr\`{\i}}, \citenamefont
  {Savasta},\ and\ \citenamefont {Nori}}]{Kockum2017}%
  \BibitemOpen
  \bibfield  {author} {\bibinfo {author} {\bibfnamefont {A.~F.}\ \bibnamefont
  {Kockum}}, \bibinfo {author} {\bibfnamefont {A.}~\bibnamefont {Miranowicz}},
  \bibinfo {author} {\bibfnamefont {V.}~\bibnamefont {Macr\`{\i}}}, \bibinfo
  {author} {\bibfnamefont {S.}~\bibnamefont {Savasta}}, \ and\ \bibinfo
  {author} {\bibfnamefont {F.}~\bibnamefont {Nori}},\ }\bibfield  {title}
  {\enquote {\bibinfo {title} {Deterministic quantum nonlinear optics with
  single atoms and virtual photons},}\ }\href {\doibase
  10.1103/PhysRevA.95.063849} {\bibfield  {journal} {\bibinfo  {journal} {Phys.
  Rev. A}\ }\textbf {\bibinfo {volume} {95}},\ \bibinfo {pages} {063849}
  (\bibinfo {year} {2017})}\BibitemShut {NoStop}%
\bibitem [{\citenamefont {Zhao}\ \emph {et~al.}(2017)\citenamefont {Zhao},
  \citenamefont {Tan}, \citenamefont {Yu}, \citenamefont {Zhu},\ and\
  \citenamefont {Yu}}]{Zhao2017}%
  \BibitemOpen
  \bibfield  {author} {\bibinfo {author} {\bibfnamefont {P.}~\bibnamefont
  {Zhao}}, \bibinfo {author} {\bibfnamefont {X.}~\bibnamefont {Tan}}, \bibinfo
  {author} {\bibfnamefont {H.}~\bibnamefont {Yu}}, \bibinfo {author}
  {\bibfnamefont {S.-L.}\ \bibnamefont {Zhu}}, \ and\ \bibinfo {author}
  {\bibfnamefont {Y.}~\bibnamefont {Yu}},\ }\bibfield  {title} {\enquote
  {\bibinfo {title} {Circuit qed with qutrits: Coupling three or more atoms via
  virtual-photon exchange},}\ }\href {\doibase 10.1103/PhysRevA.96.043833}
  {\bibfield  {journal} {\bibinfo  {journal} {Phys. Rev. A}\ }\textbf {\bibinfo
  {volume} {96}},\ \bibinfo {pages} {043833} (\bibinfo {year}
  {2017})}\BibitemShut {NoStop}%
\bibitem [{\citenamefont {Stassi}\ \emph {et~al.}(2017)\citenamefont {Stassi},
  \citenamefont {Macr\`{\i}}, \citenamefont {Kockum}, \citenamefont
  {Di~Stefano}, \citenamefont {Miranowicz}, \citenamefont {Savasta},\ and\
  \citenamefont {Nori}}]{Stassi2017}%
  \BibitemOpen
  \bibfield  {author} {\bibinfo {author} {\bibfnamefont {R.}~\bibnamefont
  {Stassi}}, \bibinfo {author} {\bibfnamefont {V.}~\bibnamefont {Macr\`{\i}}},
  \bibinfo {author} {\bibfnamefont {A.~F.}\ \bibnamefont {Kockum}}, \bibinfo
  {author} {\bibfnamefont {O.}~\bibnamefont {Di~Stefano}}, \bibinfo {author}
  {\bibfnamefont {A.}~\bibnamefont {Miranowicz}}, \bibinfo {author}
  {\bibfnamefont {S.}~\bibnamefont {Savasta}}, \ and\ \bibinfo {author}
  {\bibfnamefont {F.}~\bibnamefont {Nori}},\ }\bibfield  {title} {\enquote
  {\bibinfo {title} {Quantum nonlinear optics without photons},}\ }\href
  {\doibase 10.1103/PhysRevA.96.023818} {\bibfield  {journal} {\bibinfo
  {journal} {Phys. Rev. A}\ }\textbf {\bibinfo {volume} {96}},\ \bibinfo
  {pages} {023818} (\bibinfo {year} {2017})}\BibitemShut {NoStop}%
\bibitem [{\citenamefont {Liu}\ \emph {et~al.}(2019)\citenamefont {Liu},
  \citenamefont {Liao}, \citenamefont {Fang},\ and\ \citenamefont
  {Liu}}]{Liu2019}%
  \BibitemOpen
  \bibfield  {author} {\bibinfo {author} {\bibfnamefont {X.}~\bibnamefont
  {Liu}}, \bibinfo {author} {\bibfnamefont {Q.}~\bibnamefont {Liao}}, \bibinfo
  {author} {\bibfnamefont {G.}~\bibnamefont {Fang}}, \ and\ \bibinfo {author}
  {\bibfnamefont {S.}~\bibnamefont {Liu}},\ }\bibfield  {title} {\enquote
  {\bibinfo {title} {Dynamic generation of multi-qubit entanglement in the
  ultrastrong-coupling regime},}\ }\href {\doibase 10.1038/s41598-019-39265-4}
  {\bibfield  {journal} {\bibinfo  {journal} {Scientific Reports}\ }\textbf
  {\bibinfo {volume} {9}},\ \bibinfo {pages} {2919} (\bibinfo {year}
  {2019})}\BibitemShut {NoStop}%
\bibitem [{\citenamefont {S\'anchez Mu\~noz}\ \emph {et~al.}(2020)\citenamefont
  {S\'anchez Mu\~noz}, \citenamefont {Frisk~Kockum}, \citenamefont
  {Miranowicz},\ and\ \citenamefont {Nori}}]{Munoz2020}%
  \BibitemOpen
  \bibfield  {author} {\bibinfo {author} {\bibfnamefont {C.}~\bibnamefont
  {S\'anchez Mu\~noz}}, \bibinfo {author} {\bibfnamefont {A.}~\bibnamefont
  {Frisk~Kockum}}, \bibinfo {author} {\bibfnamefont {A.}~\bibnamefont
  {Miranowicz}}, \ and\ \bibinfo {author} {\bibfnamefont {F.}~\bibnamefont
  {Nori}},\ }\bibfield  {title} {\enquote {\bibinfo {title} {Simulating
  ultrastrong-coupling processes breaking parity conservation in
  jaynes-cummings systems},}\ }\href {\doibase 10.1103/PhysRevA.102.033716}
  {\bibfield  {journal} {\bibinfo  {journal} {Phys. Rev. A}\ }\textbf {\bibinfo
  {volume} {102}},\ \bibinfo {pages} {033716} (\bibinfo {year}
  {2020})}\BibitemShut {NoStop}%
\bibitem [{\citenamefont {You}\ and\ \citenamefont {Nori}(2011)}]{You2011}%
  \BibitemOpen
  \bibfield  {author} {\bibinfo {author} {\bibfnamefont {J.~Q.}\ \bibnamefont
  {You}}\ and\ \bibinfo {author} {\bibfnamefont {F.}~\bibnamefont {Nori}},\
  }\bibfield  {title} {\enquote {\bibinfo {title} {Atomic physics and quantum
  optics using superconducting circuits},}\ }\href {\doibase
  10.1038/nature10122} {\bibfield  {journal} {\bibinfo  {journal} {Nature}\
  }\textbf {\bibinfo {volume} {474}},\ \bibinfo {pages} {589--597} (\bibinfo
  {year} {2011})}\BibitemShut {NoStop}%
\bibitem [{\citenamefont {Devoret}\ and\ \citenamefont
  {Schoelkopf}(2013)}]{Devoret2013}%
  \BibitemOpen
  \bibfield  {author} {\bibinfo {author} {\bibfnamefont {M.~H.}\ \bibnamefont
  {Devoret}}\ and\ \bibinfo {author} {\bibfnamefont {R.~J.}\ \bibnamefont
  {Schoelkopf}},\ }\bibfield  {title} {\enquote {\bibinfo {title}
  {Superconducting circuits for quantum information: An outlook},}\ }\href
  {\doibase 10.1126/science.1231930} {\bibfield  {journal} {\bibinfo  {journal}
  {Science}\ }\textbf {\bibinfo {volume} {339}},\ \bibinfo {pages} {1169--1174}
  (\bibinfo {year} {2013})}\BibitemShut {NoStop}%
\bibitem [{\citenamefont {Kjaergaard}\ \emph {et~al.}(2020)\citenamefont
  {Kjaergaard}, \citenamefont {Schwartz}, \citenamefont {Braum\"uller},
  \citenamefont {Krantz}, \citenamefont {Wang}, \citenamefont {Gustavsson},\
  and\ \citenamefont {Oliver}}]{Kjaergaard2020}%
  \BibitemOpen
  \bibfield  {author} {\bibinfo {author} {\bibfnamefont {M.}~\bibnamefont
  {Kjaergaard}}, \bibinfo {author} {\bibfnamefont {M.~E.}\ \bibnamefont
  {Schwartz}}, \bibinfo {author} {\bibfnamefont {J.}~\bibnamefont
  {Braum\"uller}}, \bibinfo {author} {\bibfnamefont {P.}~\bibnamefont
  {Krantz}}, \bibinfo {author} {\bibfnamefont {J.~I.-J.}\ \bibnamefont {Wang}},
  \bibinfo {author} {\bibfnamefont {S.}~\bibnamefont {Gustavsson}}, \ and\
  \bibinfo {author} {\bibfnamefont {W.~D.}\ \bibnamefont {Oliver}},\ }\bibfield
   {title} {\enquote {\bibinfo {title} {Superconducting qubits: Current state
  of play},}\ }\href {\doibase 10.1146/annurev-conmatphys-031119-050605}
  {\bibfield  {journal} {\bibinfo  {journal} {Annual Review of Condensed Matter
  Physics}\ }\textbf {\bibinfo {volume} {11}},\ \bibinfo {pages} {369--395}
  (\bibinfo {year} {2020})}\BibitemShut {NoStop}%
\bibitem [{\citenamefont {Krantz}\ \emph {et~al.}(2019)\citenamefont {Krantz},
  \citenamefont {Kjaergaard}, \citenamefont {Yan}, \citenamefont {Orlando},
  \citenamefont {Gustavsson},\ and\ \citenamefont {Oliver}}]{Krantz2019}%
  \BibitemOpen
  \bibfield  {author} {\bibinfo {author} {\bibfnamefont {P.}~\bibnamefont
  {Krantz}}, \bibinfo {author} {\bibfnamefont {M.}~\bibnamefont {Kjaergaard}},
  \bibinfo {author} {\bibfnamefont {F.}~\bibnamefont {Yan}}, \bibinfo {author}
  {\bibfnamefont {T.~P.}\ \bibnamefont {Orlando}}, \bibinfo {author}
  {\bibfnamefont {S.}~\bibnamefont {Gustavsson}}, \ and\ \bibinfo {author}
  {\bibfnamefont {W.~D.}\ \bibnamefont {Oliver}},\ }\bibfield  {title}
  {\enquote {\bibinfo {title} {A quantum engineer's guide to superconducting
  qubits},}\ }\href {\doibase 10.1063/1.5089550} {\bibfield  {journal}
  {\bibinfo  {journal} {Applied Physics Reviews}\ }\textbf {\bibinfo {volume}
  {6}},\ \bibinfo {pages} {021318} (\bibinfo {year} {2019})}\BibitemShut
  {NoStop}%
\bibitem [{\citenamefont {Muthukrishnan}\ \emph {et~al.}(2004)\citenamefont
  {Muthukrishnan}, \citenamefont {Agarwal},\ and\ \citenamefont
  {Scully}}]{Muthukrishnan2004}%
  \BibitemOpen
  \bibfield  {author} {\bibinfo {author} {\bibfnamefont {A.}~\bibnamefont
  {Muthukrishnan}}, \bibinfo {author} {\bibfnamefont {G.~S.}\ \bibnamefont
  {Agarwal}}, \ and\ \bibinfo {author} {\bibfnamefont {M.~O.}\ \bibnamefont
  {Scully}},\ }\bibfield  {title} {\enquote {\bibinfo {title} {Inducing
  disallowed two-atom transitions with temporally entangled photons},}\ }\href
  {\doibase 10.1103/PhysRevLett.93.093002} {\bibfield  {journal} {\bibinfo
  {journal} {Phys. Rev. Lett.}\ }\textbf {\bibinfo {volume} {93}},\ \bibinfo
  {pages} {093002} (\bibinfo {year} {2004})}\BibitemShut {NoStop}%
\bibitem [{\citenamefont {Zheng}\ \emph {et~al.}(2013)\citenamefont {Zheng},
  \citenamefont {Saldanha}, \citenamefont {Rios~Leite},\ and\ \citenamefont
  {Fabre}}]{Zhang2013}%
  \BibitemOpen
  \bibfield  {author} {\bibinfo {author} {\bibfnamefont {Z.}~\bibnamefont
  {Zheng}}, \bibinfo {author} {\bibfnamefont {P.~L.}\ \bibnamefont {Saldanha}},
  \bibinfo {author} {\bibfnamefont {J.~R.}\ \bibnamefont {Rios~Leite}}, \ and\
  \bibinfo {author} {\bibfnamefont {C.}~\bibnamefont {Fabre}},\ }\bibfield
  {title} {\enquote {\bibinfo {title} {Two-photon--two-atom excitation by
  correlated light states},}\ }\href {\doibase 10.1103/PhysRevA.88.033822}
  {\bibfield  {journal} {\bibinfo  {journal} {Phys. Rev. A}\ }\textbf {\bibinfo
  {volume} {88}},\ \bibinfo {pages} {033822} (\bibinfo {year}
  {2013})}\BibitemShut {NoStop}%
\bibitem [{\citenamefont {Dorfman}\ \emph {et~al.}(2016)\citenamefont
  {Dorfman}, \citenamefont {Schlawin},\ and\ \citenamefont
  {Mukamel}}]{Dorfman2016}%
  \BibitemOpen
  \bibfield  {author} {\bibinfo {author} {\bibfnamefont {K.~E.}\ \bibnamefont
  {Dorfman}}, \bibinfo {author} {\bibfnamefont {F.}~\bibnamefont {Schlawin}}, \
  and\ \bibinfo {author} {\bibfnamefont {S.}~\bibnamefont {Mukamel}},\
  }\bibfield  {title} {\enquote {\bibinfo {title} {Nonlinear optical signals
  and spectroscopy with quantum light},}\ }\href {\doibase
  10.1103/RevModPhys.88.045008} {\bibfield  {journal} {\bibinfo  {journal}
  {Rev. Mod. Phys.}\ }\textbf {\bibinfo {volume} {88}},\ \bibinfo {pages}
  {045008} (\bibinfo {year} {2016})}\BibitemShut {NoStop}%
\bibitem [{\citenamefont {Ren}\ \emph {et~al.}(2020)\citenamefont {Ren},
  \citenamefont {Liu}, \citenamefont {Song}, \citenamefont {Li}, \citenamefont
  {Guo}, \citenamefont {Wang}, \citenamefont {Zheng}, \citenamefont {Agarwal},
  \citenamefont {Scully}, \citenamefont {Zhu}, \citenamefont {Wang},\ and\
  \citenamefont {Wang}}]{Ren2020}%
  \BibitemOpen
  \bibfield  {author} {\bibinfo {author} {\bibfnamefont {W.}~\bibnamefont
  {Ren}}, \bibinfo {author} {\bibfnamefont {W.}~\bibnamefont {Liu}}, \bibinfo
  {author} {\bibfnamefont {C.}~\bibnamefont {Song}}, \bibinfo {author}
  {\bibfnamefont {H.}~\bibnamefont {Li}}, \bibinfo {author} {\bibfnamefont
  {Q.}~\bibnamefont {Guo}}, \bibinfo {author} {\bibfnamefont {Z.}~\bibnamefont
  {Wang}}, \bibinfo {author} {\bibfnamefont {D.}~\bibnamefont {Zheng}},
  \bibinfo {author} {\bibfnamefont {G.~S.}\ \bibnamefont {Agarwal}}, \bibinfo
  {author} {\bibfnamefont {M.~O.}\ \bibnamefont {Scully}}, \bibinfo {author}
  {\bibfnamefont {S.-Y.}\ \bibnamefont {Zhu}}, \bibinfo {author} {\bibfnamefont
  {H.}~\bibnamefont {Wang}}, \ and\ \bibinfo {author} {\bibfnamefont {D.-W.}\
  \bibnamefont {Wang}},\ }\bibfield  {title} {\enquote {\bibinfo {title}
  {Simultaneous excitation of two noninteracting atoms with time-frequency
  correlated photon pairs in a superconducting circuit},}\ }\href {\doibase
  10.1103/PhysRevLett.125.133601} {\bibfield  {journal} {\bibinfo  {journal}
  {Phys. Rev. Lett.}\ }\textbf {\bibinfo {volume} {125}},\ \bibinfo {pages}
  {133601} (\bibinfo {year} {2020})}\BibitemShut {NoStop}%
\bibitem [{\citenamefont {Zhang}\ \emph {et~al.}(2018)\citenamefont {Zhang},
  \citenamefont {Song}, \citenamefont {Wang},\ and\ \citenamefont
  {Zheng}}]{libo2018}%
  \BibitemOpen
  \bibfield  {author} {\bibinfo {author} {\bibfnamefont {L.}~\bibnamefont
  {Zhang}}, \bibinfo {author} {\bibfnamefont {C.}~\bibnamefont {Song}},
  \bibinfo {author} {\bibfnamefont {H.}~\bibnamefont {Wang}}, \ and\ \bibinfo
  {author} {\bibfnamefont {S.-B.}\ \bibnamefont {Zheng}},\ }\bibfield  {title}
  {\enquote {\bibinfo {title} {Observation of geometric phase in a dispersively
  coupled resonator-qutrit system},}\ }\href {\doibase
  10.1088/1674-1056/27/7/070303} {\bibfield  {journal} {\bibinfo  {journal}
  {Chinese Physics B}\ }\textbf {\bibinfo {volume} {27}},\ \bibinfo {eid}
  {070303} (\bibinfo {year} {2018})}\BibitemShut {NoStop}%
\bibitem [{Sup()}]{SupMat}%
  \BibitemOpen
  \href@noop {} {}\bibinfo {note} {See Supplemental Material.}\BibitemShut
  {Stop}%
\bibitem [{\citenamefont {Song}\ \emph {et~al.}(2019)\citenamefont {Song},
  \citenamefont {Xu}, \citenamefont {Li}, \citenamefont {Zhang}, \citenamefont
  {Zhang}, \citenamefont {Liu}, \citenamefont {Guo}, \citenamefont {Wang},
  \citenamefont {Ren}, \citenamefont {Hao}, \citenamefont {Feng}, \citenamefont
  {Fan}, \citenamefont {Zheng}, \citenamefont {Wang}, \citenamefont {Wang},\
  and\ \citenamefont {Zhu}}]{Song2019}%
  \BibitemOpen
  \bibfield  {author} {\bibinfo {author} {\bibfnamefont {C.}~\bibnamefont
  {Song}}, \bibinfo {author} {\bibfnamefont {K.}~\bibnamefont {Xu}}, \bibinfo
  {author} {\bibfnamefont {H.}~\bibnamefont {Li}}, \bibinfo {author}
  {\bibfnamefont {Y.-R.}\ \bibnamefont {Zhang}}, \bibinfo {author}
  {\bibfnamefont {X.}~\bibnamefont {Zhang}}, \bibinfo {author} {\bibfnamefont
  {W.}~\bibnamefont {Liu}}, \bibinfo {author} {\bibfnamefont {Q.}~\bibnamefont
  {Guo}}, \bibinfo {author} {\bibfnamefont {Z.}~\bibnamefont {Wang}}, \bibinfo
  {author} {\bibfnamefont {W.}~\bibnamefont {Ren}}, \bibinfo {author}
  {\bibfnamefont {J.}~\bibnamefont {Hao}}, \bibinfo {author} {\bibfnamefont
  {H.}~\bibnamefont {Feng}}, \bibinfo {author} {\bibfnamefont {H.}~\bibnamefont
  {Fan}}, \bibinfo {author} {\bibfnamefont {D.}~\bibnamefont {Zheng}}, \bibinfo
  {author} {\bibfnamefont {D.-W.}\ \bibnamefont {Wang}}, \bibinfo {author}
  {\bibfnamefont {H.}~\bibnamefont {Wang}}, \ and\ \bibinfo {author}
  {\bibfnamefont {S.-Y.}\ \bibnamefont {Zhu}},\ }\bibfield  {title} {\enquote
  {\bibinfo {title} {Generation of multicomponent atomic schr{\"o}dinger cat
  states of up to 20 qubits},}\ }\href {\doibase 10.1126/science.aay0600}
  {\bibfield  {journal} {\bibinfo  {journal} {Science}\ }\textbf {\bibinfo
  {volume} {365}},\ \bibinfo {pages} {574--577} (\bibinfo {year}
  {2019})}\BibitemShut {NoStop}%
\bibitem [{\citenamefont {Guo}\ \emph {et~al.}(2021)\citenamefont {Guo},
  \citenamefont {Chen}, \citenamefont {Sun}, \citenamefont {Song},
  \citenamefont {Li}, \citenamefont {Wang}, \citenamefont {Ren}, \citenamefont
  {Dong}, \citenamefont {Zheng}, \citenamefont {Zhang}, \citenamefont
  {Mondaini}, \citenamefont {Fan},\ and\ \citenamefont {Wang}}]{Guo2021}%
  \BibitemOpen
  \bibfield  {author} {\bibinfo {author} {\bibfnamefont {Q.}~\bibnamefont
  {Guo}}, \bibinfo {author} {\bibfnamefont {C.}~\bibnamefont {Chen}}, \bibinfo
  {author} {\bibfnamefont {Z.-H.}\ \bibnamefont {Sun}}, \bibinfo {author}
  {\bibfnamefont {Z.}~\bibnamefont {Song}}, \bibinfo {author} {\bibfnamefont
  {H.}~\bibnamefont {Li}}, \bibinfo {author} {\bibfnamefont {Z.}~\bibnamefont
  {Wang}}, \bibinfo {author} {\bibfnamefont {W.}~\bibnamefont {Ren}}, \bibinfo
  {author} {\bibfnamefont {H.}~\bibnamefont {Dong}}, \bibinfo {author}
  {\bibfnamefont {D.}~\bibnamefont {Zheng}}, \bibinfo {author} {\bibfnamefont
  {Y.-R.}\ \bibnamefont {Zhang}}, \bibinfo {author} {\bibfnamefont
  {R.}~\bibnamefont {Mondaini}}, \bibinfo {author} {\bibfnamefont
  {H.}~\bibnamefont {Fan}}, \ and\ \bibinfo {author} {\bibfnamefont
  {H.}~\bibnamefont {Wang}},\ }\bibfield  {title} {\enquote {\bibinfo {title}
  {Observation of energy-resolved many-body localization},}\ }\href {\doibase
  10.1038/s41567-020-1035-1} {\bibfield  {journal} {\bibinfo  {journal} {Nat.
  Phys.}\ }\textbf {\bibinfo {volume} {17}},\ \bibinfo {pages} {234--239}
  (\bibinfo {year} {2021})}\BibitemShut {NoStop}%
\bibitem [{\citenamefont {S\o{}rensen}\ and\ \citenamefont
  {M\o{}lmer}(2000)}]{Sorensen2000}%
  \BibitemOpen
  \bibfield  {author} {\bibinfo {author} {\bibfnamefont {A.}~\bibnamefont
  {S\o{}rensen}}\ and\ \bibinfo {author} {\bibfnamefont {K.}~\bibnamefont
  {M\o{}lmer}},\ }\bibfield  {title} {\enquote {\bibinfo {title} {Entanglement
  and quantum computation with ions in thermal motion},}\ }\href {\doibase
  10.1103/PhysRevA.62.022311} {\bibfield  {journal} {\bibinfo  {journal} {Phys.
  Rev. A}\ }\textbf {\bibinfo {volume} {62}},\ \bibinfo {pages} {022311}
  (\bibinfo {year} {2000})}\BibitemShut {NoStop}%
\bibitem [{\citenamefont {Hund}(1927)}]{Hund1927}%
  \BibitemOpen
  \bibfield  {author} {\bibinfo {author} {\bibfnamefont {F.}~\bibnamefont
  {Hund}},\ }\bibfield  {title} {\enquote {\bibinfo {title} {Zur deutung der
  molekelspektren. iii.}}\ }\href {\doibase 10.1007/BF01397249} {\bibfield
  {journal} {\bibinfo  {journal} {Zeitschrift f\"ur Physik}\ }\textbf {\bibinfo
  {volume} {43}},\ \bibinfo {pages} {805--826} (\bibinfo {year}
  {1927})}\BibitemShut {NoStop}%
\bibitem [{\citenamefont {Cina}\ and\ \citenamefont {Harris}(1995)}]{Cina1995}%
  \BibitemOpen
  \bibfield  {author} {\bibinfo {author} {\bibfnamefont {J.~A.}\ \bibnamefont
  {Cina}}\ and\ \bibinfo {author} {\bibfnamefont {R.~A.}\ \bibnamefont
  {Harris}},\ }\bibfield  {title} {\enquote {\bibinfo {title} {Superpositions
  of handed wave functions},}\ }\href {\doibase 10.1126/science.267.5199.832}
  {\bibfield  {journal} {\bibinfo  {journal} {Science}\ }\textbf {\bibinfo
  {volume} {267}},\ \bibinfo {pages} {832--833} (\bibinfo {year}
  {1995})}\BibitemShut {NoStop}%
\bibitem [{\citenamefont {Mason}(1984)}]{Mason1984}%
  \BibitemOpen
  \bibfield  {author} {\bibinfo {author} {\bibfnamefont {S.~F.}\ \bibnamefont
  {Mason}},\ }\bibfield  {title} {\enquote {\bibinfo {title} {Origins of
  biomolecular handedness},}\ }\href {\doibase 10.1038/311019a0} {\bibfield
  {journal} {\bibinfo  {journal} {Nature}\ }\textbf {\bibinfo {volume} {311}},\
  \bibinfo {pages} {19--23} (\bibinfo {year} {1984})}\BibitemShut {NoStop}%
\bibitem [{\citenamefont {Darqui\'e}\ \emph {et~al.}(2010)\citenamefont
  {Darqui\'e}, \citenamefont {Stoeffler}, \citenamefont {Shelkovnikov},
  \citenamefont {Daussy}, \citenamefont {Amy-Klein}, \citenamefont
  {Chardonnet}, \citenamefont {Zrig}, \citenamefont {Guy}, \citenamefont
  {Crassous},\ and\ \citenamefont {Soulard}}]{Darqui2010}%
  \BibitemOpen
  \bibfield  {author} {\bibinfo {author} {\bibfnamefont {B.}~\bibnamefont
  {Darqui\'e}}, \bibinfo {author} {\bibfnamefont {C.}~\bibnamefont
  {Stoeffler}}, \bibinfo {author} {\bibfnamefont {A.}~\bibnamefont
  {Shelkovnikov}}, \bibinfo {author} {\bibfnamefont {C.}~\bibnamefont
  {Daussy}}, \bibinfo {author} {\bibfnamefont {A.}~\bibnamefont {Amy-Klein}},
  \bibinfo {author} {\bibfnamefont {C.}~\bibnamefont {Chardonnet}}, \bibinfo
  {author} {\bibfnamefont {S.}~\bibnamefont {Zrig}}, \bibinfo {author}
  {\bibfnamefont {L.}~\bibnamefont {Guy}}, \bibinfo {author} {\bibfnamefont
  {J.}~\bibnamefont {Crassous}}, \ and\ \bibinfo {author} {\bibfnamefont
  {P.}~\bibnamefont {Soulard}},\ }\bibfield  {title} {\enquote {\bibinfo
  {title} {Progress toward a first observation of parity violation in chiral
  molecules by high-resolution laser spectroscopy},}\ }\href@noop {} {\bibfield
   {journal} {\bibinfo  {journal} {Chirality}\ }\textbf {\bibinfo {volume}
  {22}},\ \bibinfo {pages} {870--884} (\bibinfo {year} {2010})}\BibitemShut
  {NoStop}%
\bibitem [{\citenamefont {Trost}\ and\ \citenamefont
  {Hornberger}(2009)}]{Trost2009}%
  \BibitemOpen
  \bibfield  {author} {\bibinfo {author} {\bibfnamefont {J.}~\bibnamefont
  {Trost}}\ and\ \bibinfo {author} {\bibfnamefont {K.}~\bibnamefont
  {Hornberger}},\ }\bibfield  {title} {\enquote {\bibinfo {title} {Hund's
  paradox and the collisional stabilization of chiral molecules},}\ }\href
  {\doibase 10.1103/PhysRevLett.103.023202} {\bibfield  {journal} {\bibinfo
  {journal} {Phys. Rev. Lett.}\ }\textbf {\bibinfo {volume} {103}},\ \bibinfo
  {pages} {023202} (\bibinfo {year} {2009})}\BibitemShut {NoStop}%
\bibitem [{\citenamefont {Houck}\ \emph {et~al.}(2012)\citenamefont {Houck},
  \citenamefont {T\"ureci},\ and\ \citenamefont {Koch}}]{Houck2012}%
  \BibitemOpen
  \bibfield  {author} {\bibinfo {author} {\bibfnamefont {A.~A.}\ \bibnamefont
  {Houck}}, \bibinfo {author} {\bibfnamefont {H.~E.}\ \bibnamefont {T\"ureci}},
  \ and\ \bibinfo {author} {\bibfnamefont {J.}~\bibnamefont {Koch}},\
  }\bibfield  {title} {\enquote {\bibinfo {title} {On-chip quantum simulation
  with superconducting circuits},}\ }\href {\doibase 10.1038/nphys2251}
  {\bibfield  {journal} {\bibinfo  {journal} {Nature}\ }\textbf {\bibinfo
  {volume} {8}},\ \bibinfo {pages} {292--299} (\bibinfo {year}
  {2012})}\BibitemShut {NoStop}%
\bibitem [{\citenamefont {Facchi}\ \emph {et~al.}(2011)\citenamefont {Facchi},
  \citenamefont {Florio}, \citenamefont {Pascazio},\ and\ \citenamefont
  {Pepe}}]{Facchi2011}%
  \BibitemOpen
  \bibfield  {author} {\bibinfo {author} {\bibfnamefont {P.}~\bibnamefont
  {Facchi}}, \bibinfo {author} {\bibfnamefont {G.}~\bibnamefont {Florio}},
  \bibinfo {author} {\bibfnamefont {S.}~\bibnamefont {Pascazio}}, \ and\
  \bibinfo {author} {\bibfnamefont {F.~V.}\ \bibnamefont {Pepe}},\ }\bibfield
  {title} {\enquote {\bibinfo {title} {Greenberger-horne-zeilinger states and
  few-body hamiltonians},}\ }\href {\doibase 10.1103/PhysRevLett.107.260502}
  {\bibfield  {journal} {\bibinfo  {journal} {Phys. Rev. Lett.}\ }\textbf
  {\bibinfo {volume} {107}},\ \bibinfo {pages} {260502} (\bibinfo {year}
  {2011})}\BibitemShut {NoStop}%
\bibitem [{\citenamefont {Agarwal}(2013)}]{Agarwal}%
  \BibitemOpen
  \bibfield  {author} {\bibinfo {author} {\bibfnamefont {G.~S.}\ \bibnamefont
  {Agarwal}},\ }\href@noop {} {\emph {\bibinfo {title} {Quantum Optics}}}\
  (\bibinfo  {publisher} {Cambridge University Press},\ \bibinfo {year}
  {2013})\BibitemShut {NoStop}%
\end{thebibliography}%

\end{document}